\documentclass[a4paper,11pt]{article}
\pdfoutput=1 % if your are submitting a pdflatex (i.e. if you have
\usepackage{jcappub} % for details on the use of the package, please
\usepackage[T1]{fontenc} % if needed

\usepackage{hyperref}
\usepackage{amsmath}
\usepackage{amssymb}
\usepackage{mathtools}
\usepackage{graphicx}% Include figure files
\usepackage{dcolumn}% Align table columns on decimal point
\usepackage{bm}% bold math
\usepackage{color}

\def\Rcut{\ifmmode {R_\mathrm{cut}}\else{$R_\mathrm{cut}$}\fi\xspace}%

\newcommand{\dhad}{\delta_{\mathrm{had}}}
\newcommand{\dhadz}{\delta_{\mathrm{had},0}}
\newcommand{\dhado}{\delta_{\mathrm{had},1}}
\newcommand{\dhadt}{\delta_{\mathrm{had},2}}
\newcommand{\dgamz}{\delta_{\gamma,0}}
\newcommand{\dgamo}{\delta_{\gamma,1}}
\newcommand{\dgamt}{\delta_{\gamma,2}}

\usepackage[switch,pagewise]{lineno}
\usepackage{xspace}
\usepackage{subfigure}

\def\Xmax{\ifmmode {X_\mathrm{max}}\else{$X_\mathrm{max}$}\fi\xspace}%

\title{Testing effects of Lorentz invariance violation in the propagation of astroparticles with the Pierre Auger Observatory}

\author[72]{P.~Abreu,}
\author[54,52]{M.~Aglietta,}
\author[13]{J.M.~Albury,}
\author[1]{I.~Allekotte,}
\author[70]{K.~Almeida Cheminant,}
\author[8,12]{A.~Almela,}
\author[79]{J.~Alvarez-Mu\~niz,}
\author[80]{R.~Alves Batista,}
\author[54,52]{G.A.~Anastasi,}
\author[86]{L.~Anchordoqui,}
% ...

\affiliation[1]{Centro At\'omico Bariloche and Instituto Balseiro (CNEA-UNCuyo-CONICET), San Carlos de Bariloche, Argentina}
% ...
\affiliation[8]{Instituto de Tecnolog\'\i{}as en Detecci\'on y Astropart\'\i{}culas (CNEA, CONICET, UNSAM), Buenos Aires, Argentina}
% ...
\affiliation[12]{Universidad Tecnol\'ogica Nacional -- Facultad Regional Buenos Aires, Buenos Aires, Argentina}
\affiliation[13]{University of Adelaide, Adelaide, S.A., Australia}
% ...
\affiliation[52]{INFN, Sezione di Torino, Torino, Italy}
% ...
\affiliation[54]{Osservatorio Astrofisico di Torino (INAF), Torino, Italy}
% ...
\affiliation[70]{Institute of Nuclear Physics PAN, Krakow, Poland}
% ...
\affiliation[72]{Laborat\'orio de Instrumenta\c{c}\~ao e F\'\i{}sica Experimental de Part\'\i{}culas -- LIP and Instituto Superior T\'ecnico -- IST, Universidade de Lisboa -- UL, Lisboa, Portugal}
% ...
\affiliation[79]{Instituto Galego de F\'\i{}sica de Altas Enerx\'\i{}as (IGFAE), Universidade de Santiago de Compostela, Santiago de Compostela, Spain}
\affiliation[80]{IMAPP, Radboud University Nijmegen, Nijmegen, The Netherlands}
% ...
\affiliation[86]{Department of Physics and Astronomy, Lehman College, City University of New York, Bronx, NY, USA}
% ...
\affiliation[c]{School of Physics and Astronomy, University of Leeds, Leeds, United Kingdom}
% ...

\emailAdd{auger spokespersons@fnal.gov}

\abstract{Lorentz invariance violation (LIV) is often described by dispersion relations of the form $E_i^2=m_i^2+p_i^2+\delta_{i,n} E^{2+n}$ with delta different based on particle type $i$, with energy $E$, momentum $p$ and rest mass $m$. Kinematics and energy thresholds of interactions are modified once the LIV terms become comparable to the squared masses of the particles involved. Thus, the strongest constraints on the LIV coefficients $\delta_{i,n}$ tend to come from the highest energies. At sufficiently high energies, photons produced by cosmic ray interactions as they propagate through the Universe could be subluminal and unattenuated over cosmological distances. Cosmic ray interactions can also be modified and lead to detectable fingerprints in the energy spectrum and mass composition observed on Earth. The data collected at the Pierre Auger Observatory are therefore possibly sensitive to both the electromagnetic and hadronic sectors of LIV. In this article, we explore these two sectors by comparing the energy spectrum and the composition of cosmic rays and the upper limits on the photon flux from the Pierre Auger Observatory with simulations including LIV. Constraints on LIV parameters depend strongly on the mass composition of cosmic rays at the highest energies. For the electromagnetic sector, while no constraints can be obtained in the absence of protons beyond $10^{19}$ eV, we obtain $\dgamz > -10^{-21}$, $\dgamo > -10^{-40} \ \mathrm{eV}^{-1}$ and $\dgamt > -10^{-58} \ \mathrm{eV}^{-2}$ in the case of a subdominant proton component up to $10^{20}$ eV. For the hadronic sector, we study the best description of the data as a function of LIV coefficients and we derive constraints in the hadronic sector such as $\dhadz < 10^{-19}$, $\dhado < 10^{-38} \ \mathrm{eV}^{-1}$ and $\dhadt < 10^{-57} \ \mathrm{eV}^{-2}$ at 5$\sigma$ CL.}

\begin{document}
\maketitle
\flushbottom

\section{Introduction}
\label{sec:Intro}

According to the modern physics description of nature, Lorentz invariance (LI) is a fundamental symmetry. Several tests of LI have been performed and no evidence for symmetry breaking has been found. Nevertheless, the possibility of some level of Lorentz invariance violation (LIV) at very-high energies has been proposed in the context of theories beyond the Standard Model, quantum gravity and string theory~\cite{Mattingly:2005re}.

Ultra-high energy cosmic rays (UHECR, $E>10^{18}$~eV) are thus expected to be a sensitive probe of possible LIV~\cite{Aloisio:2000cm}. Being the highest energy particles ever detected and dominantly coming from extra-galactic sources at energies $E\gtrsim 10^{19}$ eV~\cite{Aab:2017tyv}, they represent a unique opportunity to test LIV beyond the energies available in man-made accelerators. UHE messengers interact with the background photon fields and the particles reaching Earth keep information about the main features of the interactions. Relativistic kinematics is changed under LIV assumptions and, therefore, the data measured on Earth might contain imprints of LIV.

In this work, we search for LIV signatures in the data collected at the Pierre Auger Observatory, while adopting realistic scenarios for UHECR characteristics. Two independent LIV sectors are tested. The pair production of UHE photons is simulated in the extragalactic space under LIV hypotheses and the flux of secondary photons is calculated accordingly. With that, our upper limits on the UHE photon flux are used to test LIV in the electromagnetic sector. Independently, the measured energy spectrum and distributions of depth of shower maximum, \Xmax, are used together, for the first time, to set limits in the hadronic sector, as proposed in~\cite{Boncioli:2015cqa,Boncioli:2017nec,Lang:2019kA}. Dedicated simulations considering photopion production and photodisintegration of nuclei in the extragalactic propagation under LIV assumptions are compared to the data. In this work we do not take into account the possible effects of modifications of the development of the atmospheric showers due to LIV.

In section~\ref{sec:livMech}, the phenomenological approach of the LIV hypothesis used throughout this paper is explained. A description of the Pierre Auger Observatory and of the data used here are presented in section~\ref{sec:pao}. The tested UHECR scenarios are reported in section~\ref{sec:UHECRscenario}. The results for the electromagnetic and hadronic sectors are presented independently in sections~\ref{sec:photon} and~\ref{sec:nuclei}, respectively.  Discussion and conclusions are finally addressed in section~\ref{sec:disc}.

\section{Lorentz invariance violation framework}
\label{sec:livMech}

A commonly used phenomenological approach for LIV was firstly developed by Coleman and Glashow~\cite{ColemGlash1,ColemGlash2}. It reduces the LIV effects to a change in the energy dispersion relation which can be expressed as:
\begin{equation}
  \label{eq.moddisprel}
  E_i^2 - p_i^2 = m_i^2 + \epsilon_i s_i^2,
\end{equation}
where $i$ identifies the particle, $E$, $m$ and $p$ denote, respectively, the energy, mass and momentum of the particle, and $s$ can be chosen to be either the energy, $E$, or the momentum, $p$, this choice being irrelevant for ultra-relativistic particles. We consider $s=E$ in the following. The extra term in the energy dispersion relation thus modifies the propagation and kinematics of each particle type. The experimental fact that LI is preserved at low energies, or in other words that LIV could only be significant at high energies, can be expressed by taking $\epsilon$ as function of $E$ and expanding it in a polynomial series, so that
\begin{equation}
\label{eq.moddisprel3}
E_i^2 - p_i^2 = m_i^2 + \sum^N_{n=0}\delta_{i,n} E_i^{2+n} \; ,
\end{equation}
where $n$ is the order of the perturbation. The parameters $\delta_{i,n}$ are in principle independent for each particle species and define the energy scale associated with the violation. We note that because LIV is often associated with quantum gravity theories, an expansion in terms of the Planck scale is sometimes introduced as
\begin{equation}
\label{eq.delta}
\delta_{i,n} = \frac{\eta_{i,n}}{M_{\mathrm{Pl}}^n} \; ,
\end{equation}
where $M_{\mathrm{Pl}} \approx 1.22 \times 10^{19} \ {\rm GeV}/c^2$ is the Planck mass.

If we assume that only the lowest-order non-vanishing term has a non-negligible effect, it is possible to set individual limits on $\delta_{i,n}$.

%==============================================
\section{The Pierre Auger Observatory and the datasets}
\label{sec:pao}

The Pierre Auger Observatory~\cite{pao1} is the largest UHECR detector currently in operation. Located in the southern hemisphere in western Argentina, just northeast of the town of Malargüe (69$^\circ$W, 35$^\circ$S, 1400 m a.s.l.), it covers an area of 3000 km$^2$ with a Surface Detector array (SD)~\cite{paoSD} overlooked by a Fluorescence Detector (FD)~\cite{paoFD}. The SD consists of 1660 water-Cherenkov detectors arranged in a triangular grid operating with a nearly 100\% duty cycle. Each SD station detects at ground level the secondary particles of the extensive air shower (EAS) produced by the primary UHECR interacting in the atmosphere. The FD consists of a set of telescopes that measure the UV fluorescence light from nitrogen molecules excited by the EAS particles along their path in the atmosphere. FD operations are limited to clear moonless nights, resulting in a duty cycle of $\sim$15\%~\cite{paoFDExp}.

The combination of the information from both techniques results in a quasi-calorimetric determination of the energy scale, a geometric direction reconstruction and an estimator of the primary particle mass. Details of the reconstruction techniques and their efficiency can be found in~\cite{pao1}. The determination of the features of the energy spectrum measured by the Pierre Auger Observatory reached unprecedented precision \cite{Aab:2020gxe,PierreAuger:2020kuy}, achieved with a continuous and stable operation of the detectors combined with the hybrid energy calibration. In this analysis, we use the energy spectrum measured using an accumulated exposure of 60,400 $\pm$ 1,800 $\mathrm{km^2}$ sr yr, obtained by the combination of complementary data sets above $3 \times 10^{18}$ eV based on 215,030 events recorded with zenith angles below 60$^\circ$~\cite{bib:auger:spectrum:icrc2019}. The events were measured by the SD between 1 January 2004 and 31 August 2018. A subset events with energy above $3\times10^{18}$ eV were collected between 1 January 2004 and 31 December 2017 simultaneously by the SD and FD detectors. This subset is used to calibrate the energy reconstruction of the events.

The atmospheric depth at which the air shower peaks in the rate of energy deposit, \Xmax, is the most accurate available parameter with proven correlation with the mass of the primary particle. The FD measures the fluorescence light produced in the development of the shower in the atmosphere from which the \Xmax can be reconstructed. In this paper, we use \Xmax distributions of the events measured from 1 December 2004 to 31 December 2017 with energy above $10^{17.8}$ eV and zenith angle below 65$^\circ$~\cite{Aab:2019ogu}. The 35425 events surviving the analysis and selection procedure~\cite{Aab:2014kda} are sampled into the \Xmax distribution for each of 19 energy bins [$10^{17.8},10^{17.9}$) eV, $\ldots$, [$10^{19.5},10^{19.6}$) eV, and [$10^{19.6}$ eV,$+\infty$).

No photon has yet been conclusively detected by the Pierre Auger Observatory, leading to the most restrictive upper limits on the photon flux at the highest energies~\cite{Aab:2016agp,2019ICRC...36..398R,2021ICRC...373}. Multivariate analyses based on SD and FD data was perfomed in the all events with zenith angle between 30$^\circ$ and 60$^\circ$ measured from 1 January 2004 to 30 June 2018. Since no photon-like event was found with an exposure of 40,000 km$^2$ sr yr, the null result was transformed into flux upper limits for energies above $10^{18}$ eV

\section{The UHECR scenario}
\label{sec:UHECRscenario}

In the present work, we test the sensitivity of the data collected by the Pierre Auger Observatory to violations of LI under specified UHECR scenarios. We consider isotropically distributed sources with $(1+z)^m$ cosmological evolution emitting an energy spectrum given by
\begin{equation}
  \label{eq.accspectrum}
  \frac{dN_A}{dE} = J_0 f_{A} \left( \frac{E}{10^{18}~\mathrm{eV}} \right) ^{-\Gamma} \times
  \begin{cases}
    1, & \mathrm{for } \ R < \Rcut \\
    \exp{ (1-R/\Rcut) }, & \mathrm{for } \ R \ge \Rcut
  \end{cases},
\end{equation}
where $z$ is the redshift, $\Gamma$ is the spectral index at the injection, $\Rcut$ is the cutoff rigidity, $f_{A}$ is the fraction of nuclei with mass $A$, and $J_0$ is the normalization factor of the flux which enters into the computation of the total emissivity defined as $\mathcal{L}_0=\sum_A\int^{+\infty}_{E_{\mathrm{min}}} E Q_A(E) \, dE$, where $Q_A(E) \, dE$ is the number of nuclei with mass A injected per unit energy, volume and time.

UHECRs interact with photon backgrounds on the way to Earth. At these energies, the most important photon backgrounds are the cosmic microwave background (CMB) and the extra-galactic background light (EBL). Given the uncertainties in the models, we included two EBL distributions and two photo-nuclear cross section models among the UHECR propagation models used in this study, as used also in \cite{Aab:2016zth} and described in \cite{AlvesBatista:2015jem}. The comparison of events arriving at Earth with \Xmax data is only possible if the arriving mass composition is transformed into an \Xmax distribution with the use of EAS simulations in which the most relevant alternative is the hadronic interaction model. We used Gumbel functions~\cite{DeDomenico:2013wwa} to parametrize the \Xmax distribution. Two hadronic interaction models were considered.

Finally, a UHECR scenario is here defined by: a) the set of parameters from the source: $\Gamma$, $\Rcut$, $J_0$, $f_{A}$ and $m$; b) the propagation model: the EBL and photo-nuclear cross sections and; c) the hadronic interaction model used to convert the arriving composition into an \Xmax distribution. Table~\ref{tab:prop:comb} summarizes the UHECR scenarios used in the following sections.

\begin{table}[ht]
  \centering
  \begin{tabular}{c|c|c|c}
    %\multicolumn{4}{c}{Models used in each UHECR-source scenario} \\
    \hline \hline
    Scenario & Cross sections & EBL & Hadronic interactions \\  \hline
    SPGE & PSB~\cite{Puget:1976nz,Stecker:1998ib} & Gilmore~\cite{Gilmore:2011ks} & EPOS-LHC~\cite{Pierog:2013ria} \\
    STGE & Talys~\cite{Koning:2007} & Dominguez~\cite{Dominguez:2010bv} & EPOS-LHC \\
    SPDE & PSB & Gilmore & EPOS-LHC \\
    SPGS & PSB & Gilmore & Sibyll 2.3c~\cite{Engel:2019dsg} \\ \hline \hline
  \end{tabular}
  \caption{Definition of the acronyms for the combinations of propagation and hadronic interaction models used in this work. We show the options for a) photo-nuclear cross sections; b) extra-galactic background light (EBL) distribution; and c) hadronic interaction model used in this work. The first S in the acronym stands for the software SimProp v2.4~\cite{Aloisio:2017iyh}.}
  \label{tab:prop:comb}
\end{table}

%==============================================
\section{LIV limits on the electromagnetic sector}
\label{sec:photon}

In this section, we investigate LIV signatures in the electromagnetic sector.
The interaction of UHECRs with the background radiation produces pions, as first studied by Greisen\cite{Greisen:1966jv}, Zatzepin and Kuzmin~\cite{Zatsepin:1966jv} (GZK). The neutral pions decay into UHE photons shortly after, which we refer to as GZK photons and are absorbed during propagation due to electron-positron pair production with background photon fields~\cite{DeAngelis:2013jna}. As explained below, we simulate GZK photons within a given UHECR scenario under LIV and LI assumptions and compare the predicted flux arriving at Earth with the upper limits on the photons flux measured by the Pierre Auger Observatory.

%==============================================
\subsection{Simulation of the propagation of GZK photons under LIV assumption}

\begin{figure}
    \centering
    \includegraphics[width=0.6\textwidth]{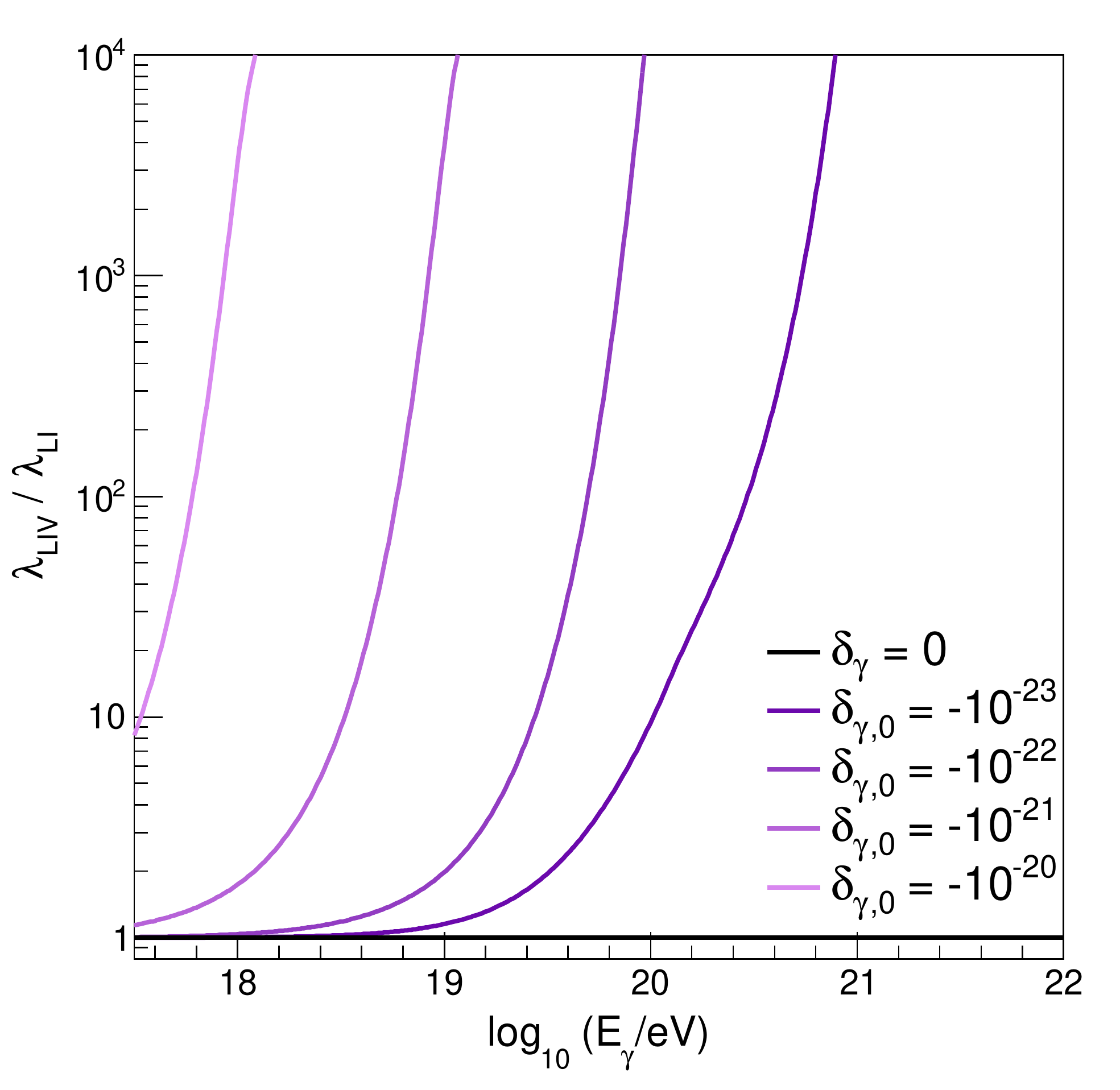}
    \caption{Ratio of the mean free path of pair production as a function of energy when LIV and LI are considered. The black line represents the LI case, while the shades of purple represent different LIV coefficients.}
    \label{fig:photonmeanfreepath}
\end{figure}

If LIV is present in the photon sector, the kinematics of pair production is changed. Following the calculations from~\cite{Lang:2017wpe}, we obtain the mean free path for this interaction considering subluminal LIV ($\delta_\gamma < 0$). Figure~\ref{fig:photonmeanfreepath} shows the ratio between the LIV and LI mean free paths for different LIV coefficients corresponding to the LIV order $n=0$. The main result is a significant increase of the mean free path above a critical energy that depends on $\dgamz$. Comparable results are found for $n=1$ and $n=2$. As a consequence, it is expected that if subluminal LIV is present, fewer UHE photons will be absorbed and, consequently, the arriving flux of these particles will be larger. As first proposed in~\cite{Galaverni:2007tq,Galaverni:2008yj} and later developed in~\cite{Lang:2017wpe}, for some UHECR scenarios the predicted flux is larger than the upper limits imposed by the Pierre Auger Observatory, resulting in limits on the LIV coefficients. If superluminal LIV ($\delta_\gamma > 0$) were considered, the mean free path would decrease, leading to a weaker flux arriving on Earth. Therefore, the upper limits on flux are insensitive to superluminal LIV in the photon sector.

The LIV-modified mean free path was implemented in the software packages CRPropa3/EleCa~\cite{Batista:2016yrx,SETTIMO2013279} in order to obtain the arriving GZK photon flux. We simulated UHECRs with energies ranging from $10^{18.7}$~eV to $10^{21}$~eV for five representative primaries (H, He, N, Si and Fe). Sources were considered homogenously distributed ($m=0$) going up to 9500~Mpc. The energy spectrum of UHECRs arriving on Earth was normalized at $E=10^{18.75}$~eV. Different LIV coefficients with $\log_{10}$ steps of 1 were considered for each order, $n$. The two limiting scenarios were also considered, LI ($\dhad=0$) and the case in which no interactions happen at any energy, here noted as $\dhad \rightarrow -\infty$.

%==============================================
\subsection{Results with the reference UHECR scenario}

First, we consider the reference SPGE propagation model and the source parameters which best describes the energy spectrum and \Xmax distributions measured by the Pierre Auger Observatory under LI assumption for that scenario: $\Gamma = -1.3$, $\log_{10}( \Rcut/V ) = 18.22$, $f_{\mathrm{H}} = 13\%$, $f_{\mathrm{He}} = 43\%$, $f_\mathrm{N} = 28\%$, $f_{\mathrm{Si}} = 14\%$, $f_{\mathrm{Fe}}=0.08\%$ and $m=0$, as found in \cite{Aab:2016zth}. Using this UHECR scenario, we simulated the GZK photons arriving at Earth under LIV assumption.

\begin{figure}
    \centering
    \includegraphics[width=0.6\textwidth]{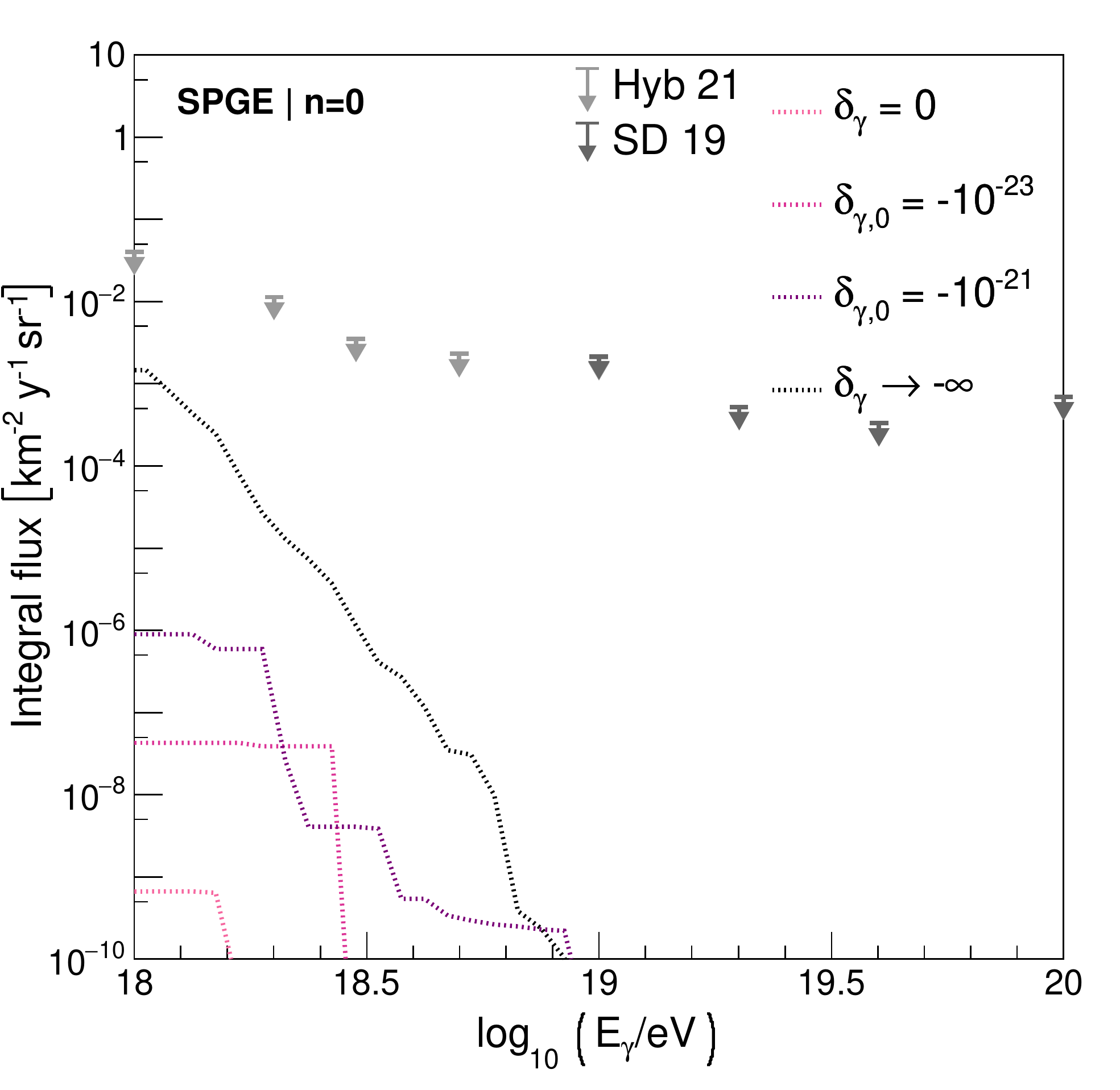}
    \caption{Simulated integral flux of GZK photons as a function of the energy for the SPGE propagation model. The arrows show the upper limits on the flux imposed by the Pierre Auger Observatory~\cite{2019ICRC...36..398R,2021ICRC...373}.}
    \label{fig:spectramixed}
\end{figure}

Figure~\ref{fig:spectramixed} shows the simulated integral fluxes of GZK photons within the SPGE propagation model and LIV order $n=0$ for several LIV coefficients ($\dgamz$), including the two limiting scenarios. Even for the most extreme LIV case ($\dgamz \rightarrow - \infty$), the simulated GZK photon flux is smaller than the measured upper limits on the photon flux and, thus, this data is insensitive to subluminal LIV in the electromagnetic sector. The same conclusion is found for all combinations of propagation and hadronic interaction models in Table~\ref{tab:prop:comb} and for LIV orders $n=1$ and $n=2$. Considering positive source evolution ($m>0$) would slightly increase the simulated GZK photon flux but not enough to reach the measured upper limits on the photon flux.

%==============================================
\subsection{Results with an alternative UHECR scenario}

As shown in~\cite{Lang:2017wpe}, under LIV hypothesis, the flux of GZK photons is expected to be more intense if a non-negligible proton component is present. In Ref.~\cite{Muzio:2019leu} it has been shown that the energy spectrum and composition data measured by the Pierre Auger Observatory can be described by models including a subdominant proton component for energies above $10^{19}$~eV. Another possibility which could contribute to the increase of GZK photons, would be to consider an additional proton component in the energy range below the ankle, together with an additional Galactic or extragalactic component, as done in \cite{Guido}. We consider only in this subsection an alternative scenario with the proton component above $10^{19}$~eV, from Ref.~\cite{Muzio:2019leu}, that best describes the energy spectrum and \Xmax distributions measured by the Pierre Auger Observatory. The integral fraction of protons for $E>10^{18.75}$~eV for this scenario is $31\%$. For more details regarding the ejected spectra, refer to Ref.~\cite{Muzio:2019leu}. The Gilmore EBL model and Talys cross-section were used. The EPOS-LHC hadronic interaction model was used to convert the arriving composition into \Xmax distributions. Using this alternative scenario, we simulated the propagation of UHECR and GZK photons in order to obtain the flux of GZK photons arriving at Earth under a LIV assumption.

\begin{figure}
    \centering
    \includegraphics[width=0.495\textwidth]{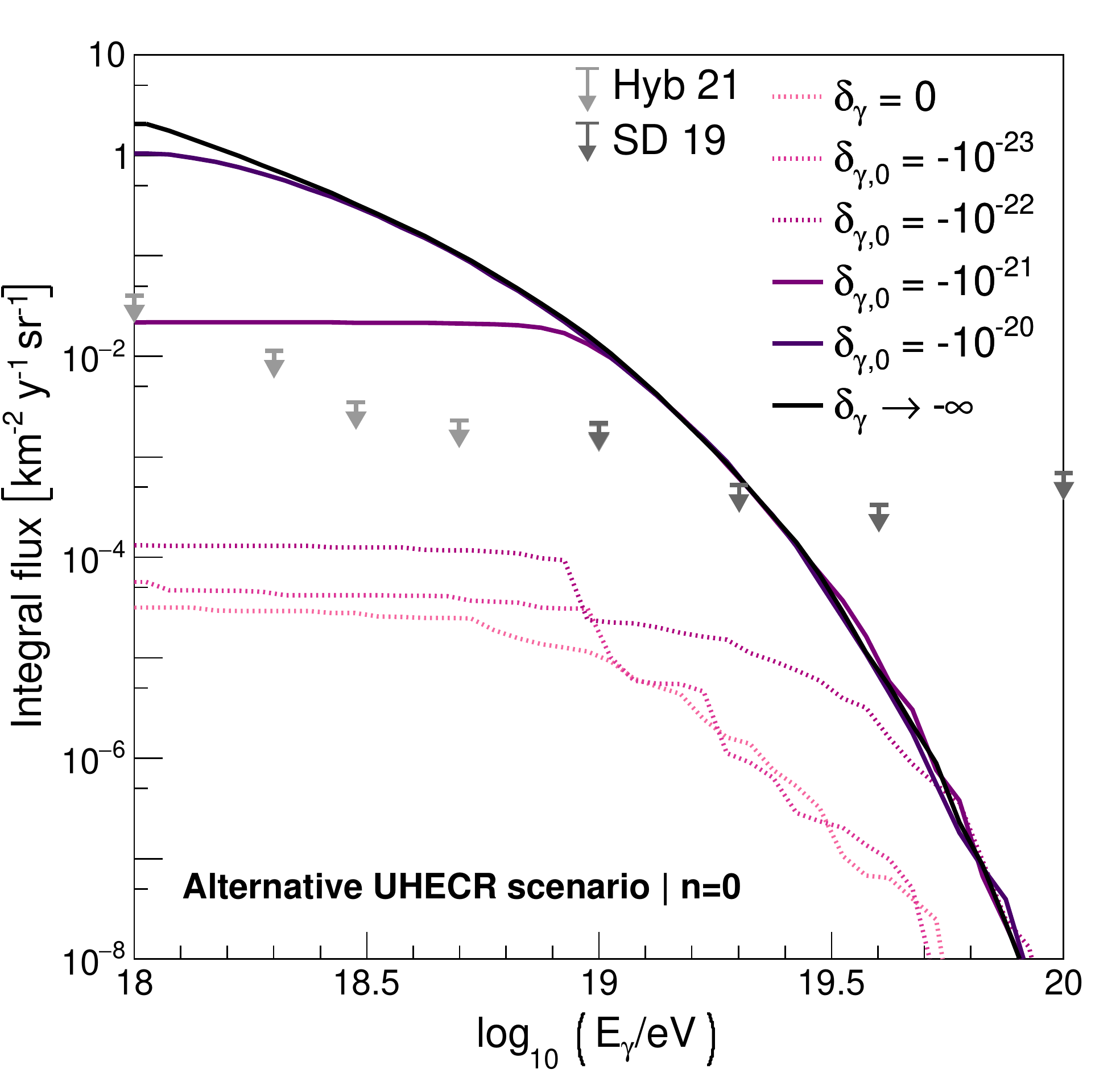}
    \includegraphics[width=0.495\textwidth]{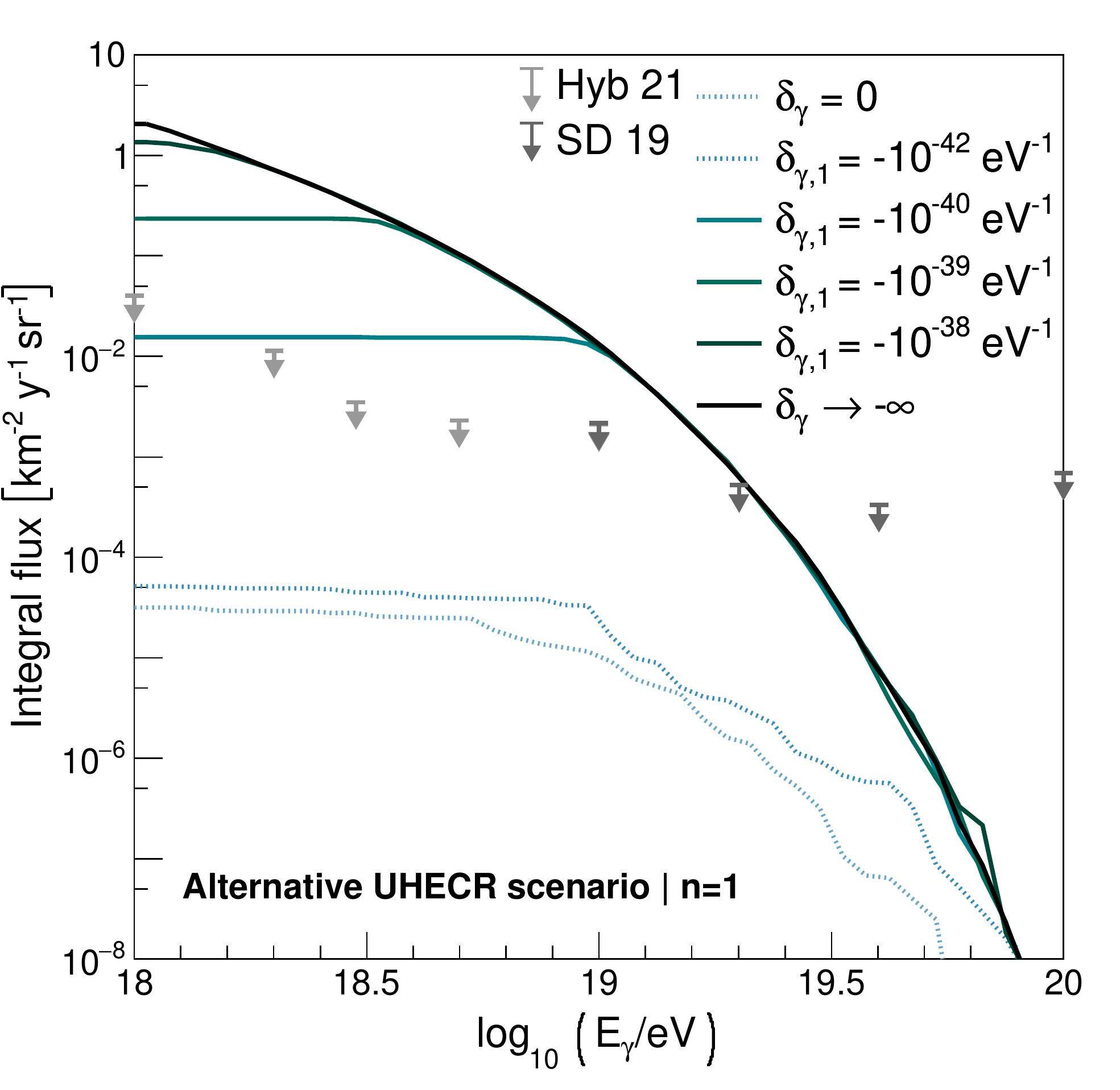}
    \includegraphics[width=0.495\textwidth]{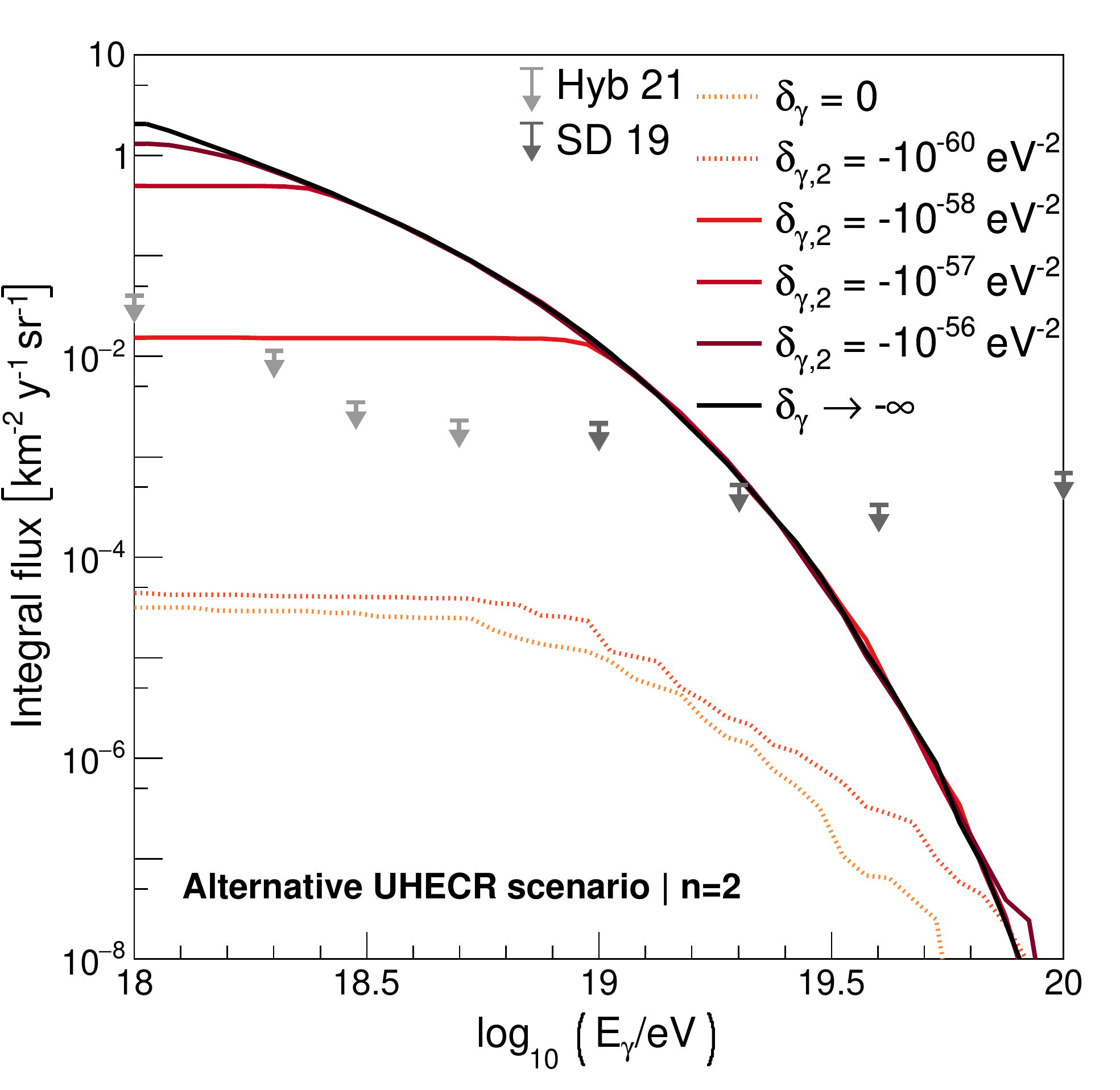}
    \caption{Simulated integral flux of GZK photons as a function of the energy for an alternative scenario with subdominant proton component~\cite{Muzio:2019leu}. Continuous lines show the rejected LIV scenarios. The arrows show the flux determined by analysis of the Pierre Auger Observatory data~\cite{2019ICRC...36..398R,2021ICRC...373}.}
    \label{fig:subdominant}
\end{figure}

Figure~\ref{fig:subdominant} shows the simulated GZK photon flux arriving at Earth under this alternative scenario for LIV orders $n$=0, 1 and 2 compared to the measured upper limits on the photon flux from the Pierre Auger Observatory. For some LIV coefficients, the simulated GZK photon flux is higher than the upper limits. Direct comparison of the simulation to the data imposes limits of $\dgamz > -10^{-21}$, $\dgamo > -10^{-40} \ \mathrm{eV}^{-1}$ and $\dgamt > -10^{-58} \ \mathrm{eV}^{-2}$.

%==============================================
\section{LIV limits on the hadronic sector}
\label{sec:nuclei}

In this section, we investigate LIV signatures in the hadronic sector. During propagation, UHECRs interact with the background radiation and the kinematics of the dominant energy losses, photopion production and photodisintegration, are modified under LIV assumptions. We consider only positive LIV coefficients ($\dhad > 0$). Less significant changes are found if negative coefficients ($\dhad < 0$) are considered as the photopion production and the photo-disintegration  cross-sections prevent a significant decrease in the attenuation length, even if the energy threshold decreases significantly.

In this section, instead of assuming an \emph{a priori} scenario, we fit the source parameters and the LIV coefficients to the energy spectrum and composition from the Pierre Auger Observatory. The latest works~\cite{Bi:2008yx,Scully:2008jp,Maccione:2009ju} considering effects of LIV in photopion production and photodisintegration with the aim of fitting UHECR data consider a simple pure-proton composition which is largely disfavored by the Pierre Auger Observatory data. Indeed, current data from the Pierre Auger Observatory exclude at 6.5$\sigma$ a pure proton or a mixed proton and helium composition at the spectral ankle region $(3 \mathrm{-} 5)\times10^{18}$ eV and is compatible with a near zero proton fraction just above $10^{19}$ eV with a possible resurgence limited to a maximum of about 10\% of the total flux above $10^{19.3}$~eV~\cite{Aab:2014aea}. As defined below, we allow the nuclei fractions, $f_A$, to vary in the fitting procedure as well as the LIV coefficients. The different combinations of propagation and hadronic interaction models from Table~\ref{tab:prop:comb} are used.

We define the LIV coefficient for hadrons in such way that $\dhad \coloneqq \delta_p = \delta_\pi/2 = A^{n} \delta_A$, where $\delta_p$, $\delta_\pi$ and $\delta_A$ are the coefficients for protons, pions and nuclei of atomic mass $A$, respectively and $n$ is the LIV order. As discussed in~\cite{Scully:2008jp}, the effects of LIV on photopion production depend on $\delta_\pi - \delta_p$, which justifies the use of the relation $\delta_p = \delta_\pi/2$ in order to have an effect of the same order in each of the interactions. The relation between $\delta_p$ and $\delta_A$, on the other hand, is supported by the superposition model
\begin{eqnarray}
  \centering
  E_{A}^{2} & = & m_A^2 + p_A^2 + \sum_{n=0}^{\infty} \delta_{A,n} E_{A}^{(n+2)} \implies \\
  A^2 E_p^2 & = & A^2 m_p^2 + A^2 p_A^2 + \sum_{n=0}^{\infty} A^{(n+2)} \delta_{A,n} E_p^{(n+2)} \implies \\
  \delta_{p,n} & = & A^{n} \delta_{A,n} \; .
\end{eqnarray}
These dispersion relations will be used to implement the modifications to the kinematics of the interactions in simulations including LIV.
%==============================================
\subsection{Simulation of the propagation of UHECR under LIV assumptions}

\begin{figure}
  \centering
  \includegraphics[width=0.6\textwidth]{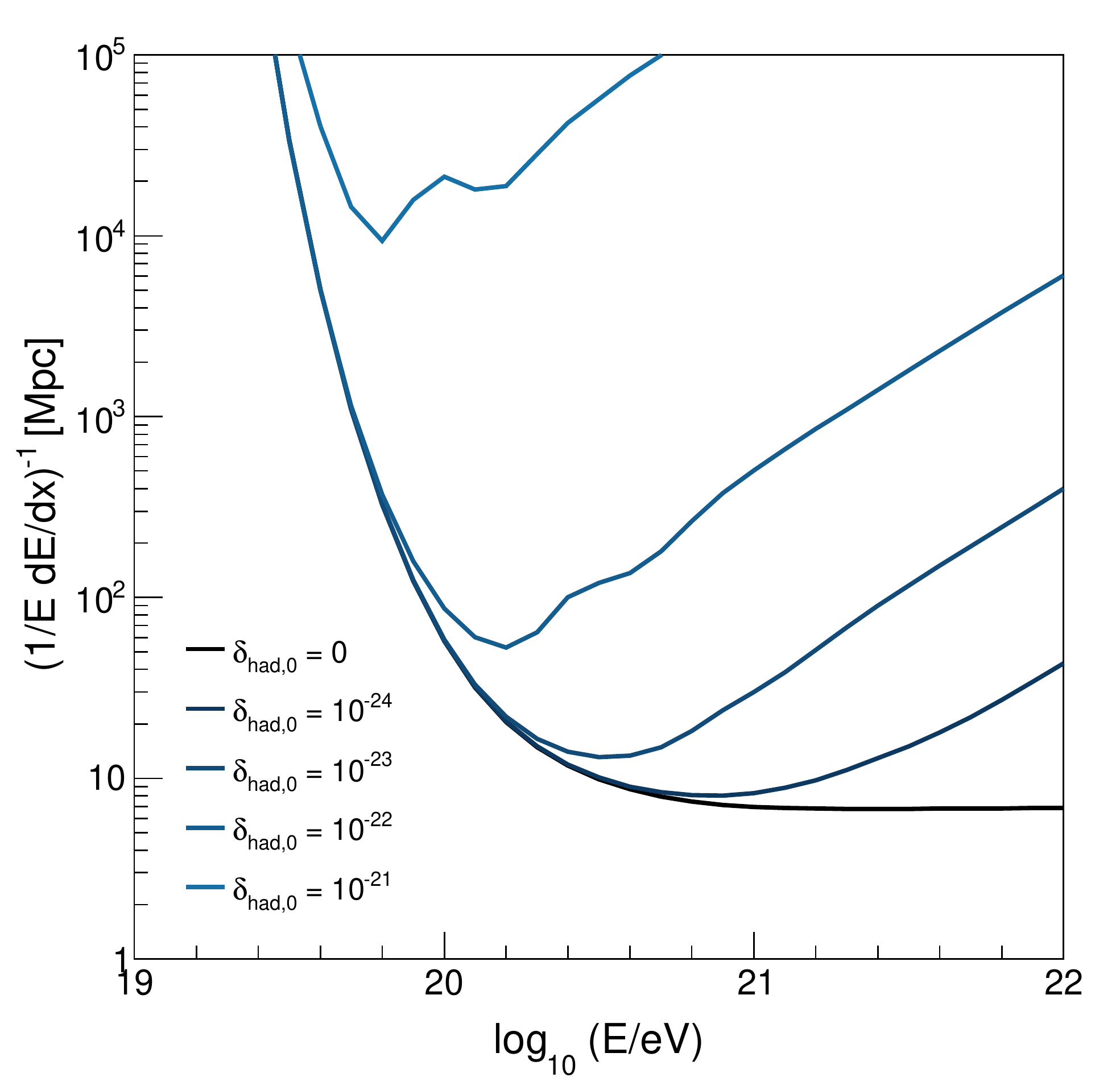}
  \caption{Attenuation length for photopion production as a function of energy for different LIV coefficients. The black line represents LI $\dhadz=0$ and lines in shades of blue represent different LIV coefficients. Strong LIV with $\dhad > 10^{-21}$ results in attenuation lengths larger than $10^{5}$~Mpc for all the considered energies and, thus, are not shown in the plot.}
  \label{fig:pionproduction}
\end{figure}

The dominant energy loss for protons at the highest energies is the photopion production ($p + \gamma \rightarrow p + \pi$) for which the attenuation length under LIV assumption was calculated based on~\cite{ThesisLang}. Figure~\ref{fig:pionproduction} shows the attenuation length as a function of energy for different values of $\dhadz$. LI is represented by $\dhadz = 0$. Above a critical energy, which depends on the value of $\dhadz$, the attenuation length is significantly larger for LIV than for the LI assumption. Above this critical energy, if $\dhadz > 0$ (LIV) the number of interactions during propagation is reduced and, consequently, the cosmic rays can travel farther than they would do under the LI hypothesis.

\begin{figure}
  \centering
  \includegraphics[width=0.48\textwidth]{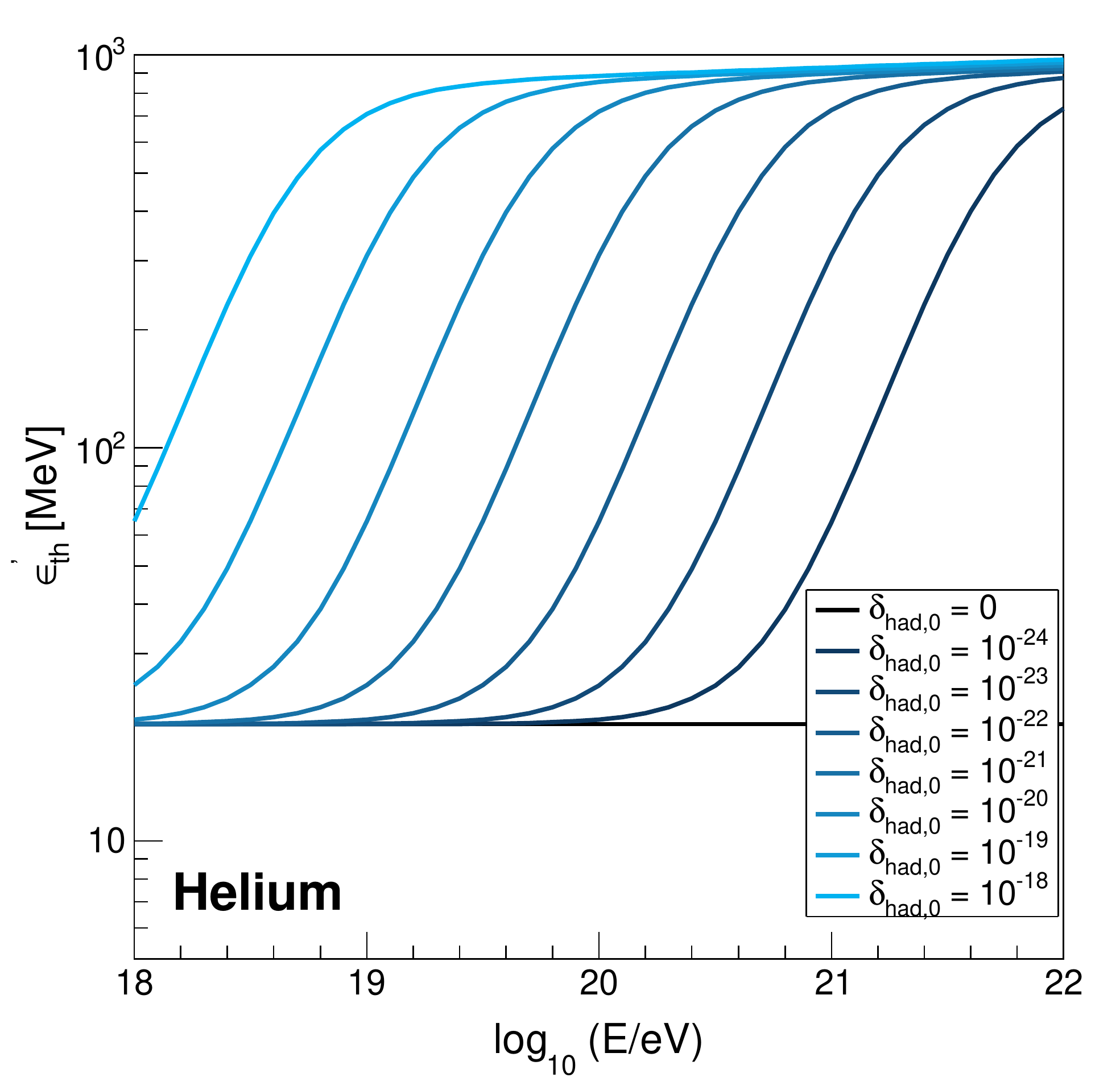}
  \includegraphics[width=0.48\textwidth]{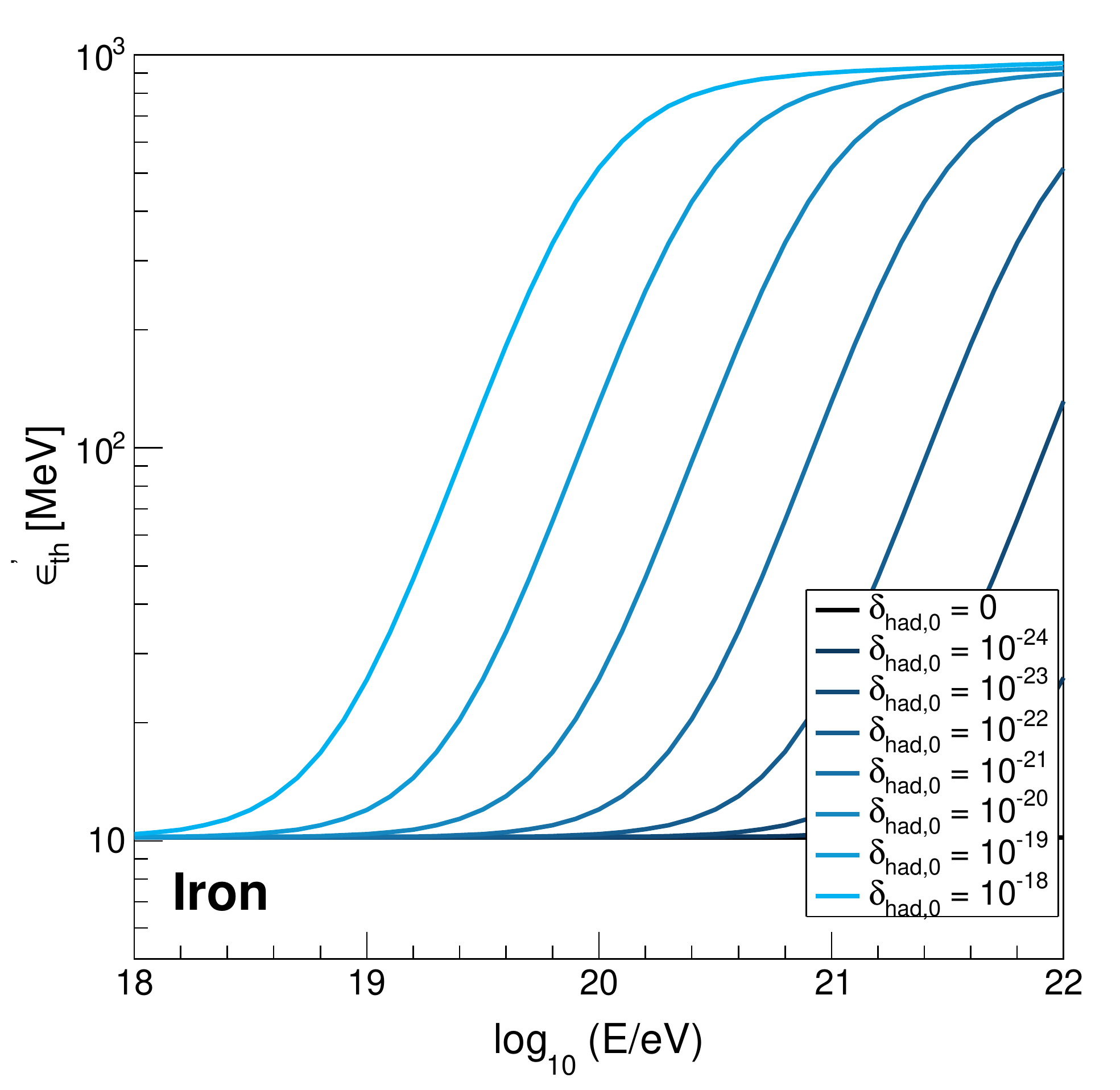}
  \caption{Energy threshold in the nucleus reference frame for photodisintegration as a function of energy for different LIV coefficients. The black lines represent the LI scenario while the shades of blue represent different LIV coefficients. The left and right panels show the results for a nucleus of helium and iron, respectively.}
  \label{fig:photodis}
\end{figure}

For nuclei, on the other hand, the leading energy loss is the photodisintegration (${^{A}N} + \gamma \rightarrow (p \; \mathrm{or} \; n) + {^{A-1}N}$) with the emission of one nucleon. Multinucleon emission is less relevant and therefore not taken into account together with the LIV modifications in the propagation. Figure~\ref{fig:photodis} shows the photon energy threshold in the nuclei reference frame for different values of $\dhadz$ and for two different nuclei (helium and iron), obtained following the calculations from~\cite{ThesisLang}. Under the LIV assumption the energy threshold increases abruptly above an energy that depends on $\dhadz$. Comparable results are found for the other nuclei and LIV orders, $n$. In a similar way to the photopion production, under the LIV assumption the UHECRs would interact less and, thus, travel farther than under the LI hypothesis.

Both the LIV modified attenuation length for pion production and the LIV modified energy threshold for photodisintegration were implemented in SimProp v2.4~\cite{Aloisio:2017iyh}. Simulations were performed for five representative nucleus (H, He, N, Si and Fe) divided into seven redshift intervals: [0,0.01), [0.01,0.05), [0.05,0.1), [0.1,0.2), [0.2,0.3), [0.3,0.5) and [0.5,2.5). The energy range of the simulation was $17.5 \le \log_{10}(E/\mathrm{eV}) \le 22$. We simulated $10^5$ events for each nucleus and redshift interval resulting in a total of $3.5 \times 10^6$ events.

%==============================================
\subsection{Results of the combined fit considering LIV}
\label{sec:nucleimixed}

\begin{figure}[h!]
    \centering
    \includegraphics[width=0.48\textwidth]{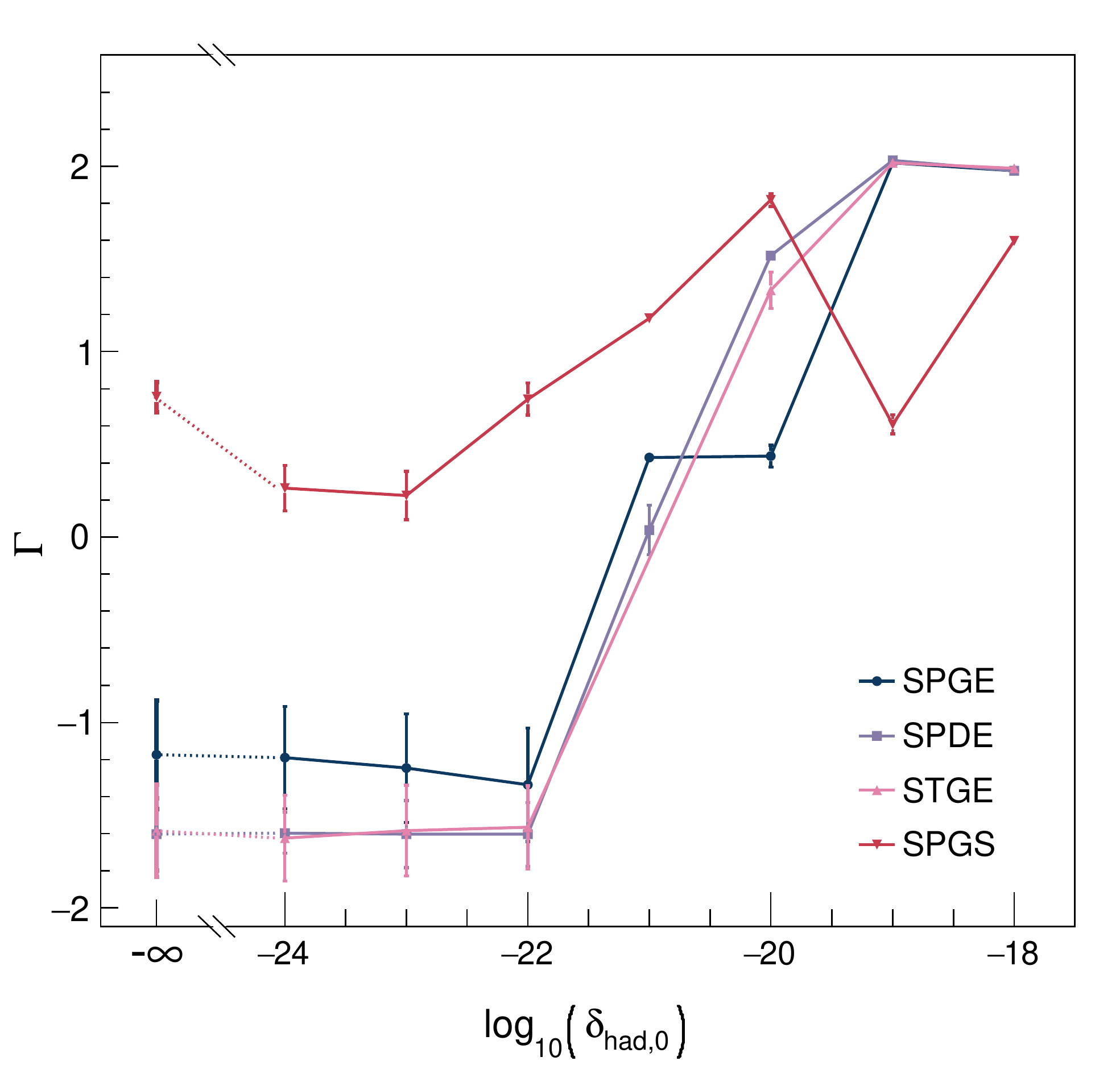}
    \includegraphics[width=0.48\textwidth]{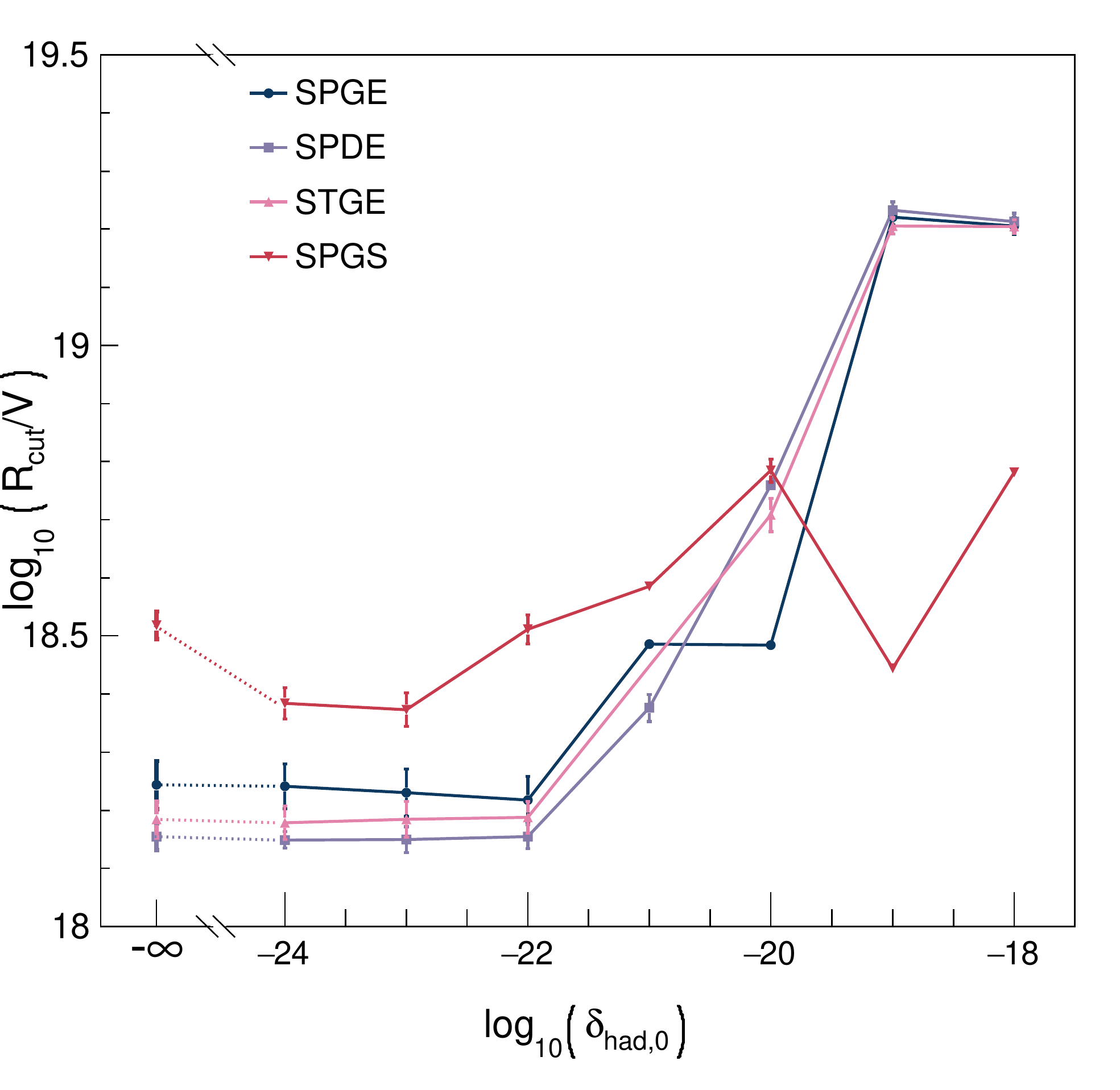}
    \includegraphics[width=0.48\textwidth]{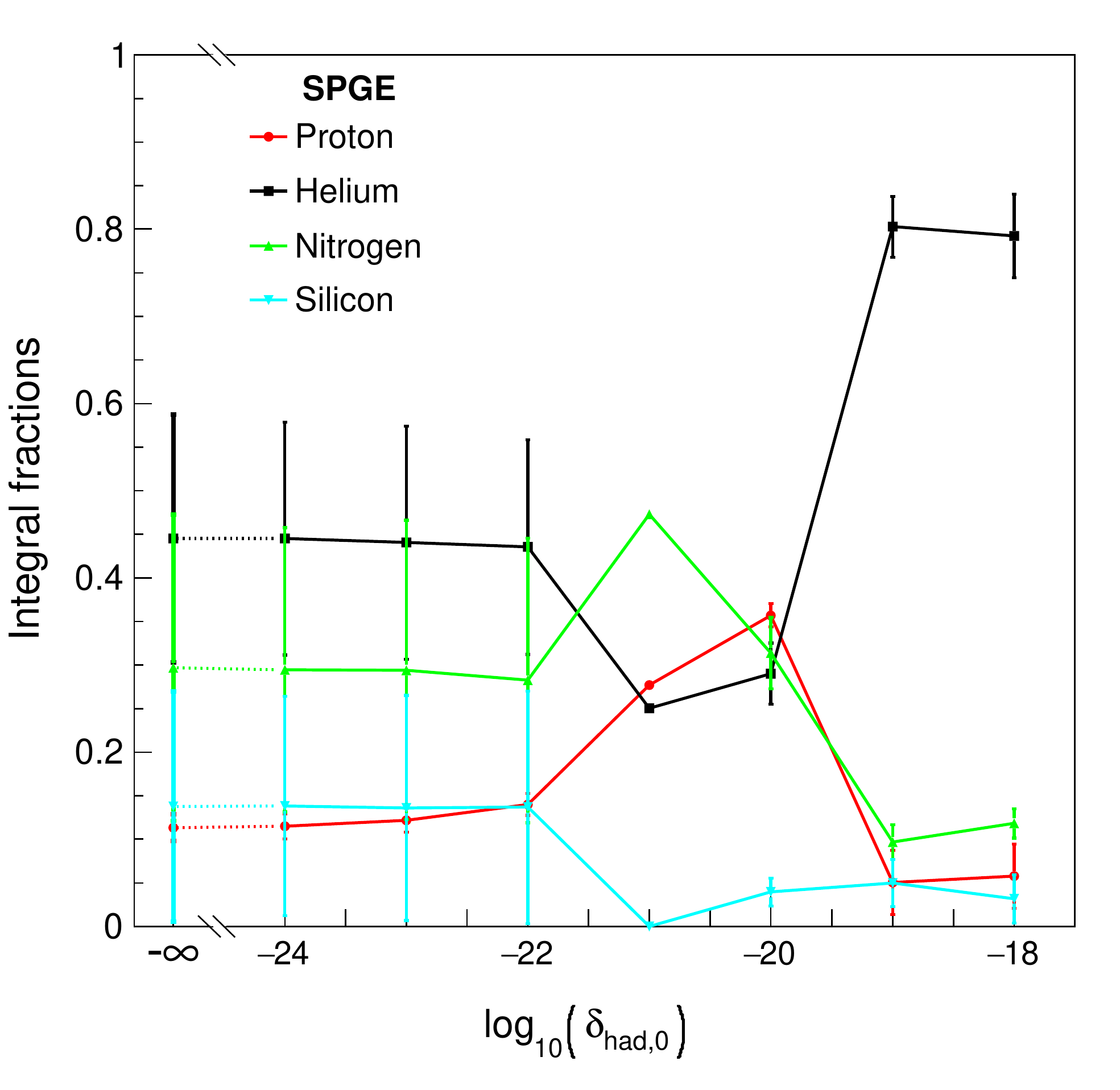}
    \caption{Evolution of the fit parameters with respect to the LIV coefficient, $\dhadz$. The panels show the spectral index, rigidity cutoff and integral fractions of each nuclei at the source, respectively. In the first two panels, each line represents a propagation model. In the last panel, each line is the percentage of each nucleus summed in the entire energy range for the SPGE model. }
    \label{fig:fitparameters}
\end{figure}

The simulated energy spectrum and composition arriving at Earth are used to fit to the data of the Pierre Auger Observatory for energies above $10^{18.7}$~eV, using Eq.~\ref{eq.accspectrum} for weighing the UHECR spectrum at the source. The fit procedure follows the explanation in~\cite{Aab:2016zth}. Within each UHECR scenario, the free parameters of the fit are: a) the nuclei fractions, $f_A$, b) the index of the energy spectrum, $\Gamma$, c) the maximum rigidity, $\Rcut$, d) the normalization factor of the flux, $J_0$ and e) the LIV coefficient, $\dhadz$. The cosmological evolution of the sources was fixed to $m=0$. For each value of $\dhadz$ ranging from $10^{-24}$ to $10^{-18}$ in $\log_{10}$ steps of 1 and for the LI case ($\dhadz = 0$) a log-likelihood fit was done searching for the combination of the parameters which best describes the data. Figure~\ref{fig:fitparameters} shows the evolution of the best fit parameters as a function of $\dhadz$.

Figure~\ref{fig:deviance} shows the total deviance, $D$, and the deviance relative ($D_{\mathrm{LIV}}-D_{\mathrm{LI}}$) to the LI case. Details on how the deviance is defined and calculated can be found in~\cite{Aab:2016zth}. The deviance becomes too large for strong LIV (large values of $\dhadz$), and, thus, limits on $\dhadz$ can be imposed with a confidence level (CL) given by $\sigma = \sqrt{ D_{\mathrm{LIV}}-D_{\mathrm{LI}}}$. For all UHECR scenarios composed of options shown in Table~\ref{tab:prop:comb}, the data imply that $\dhadz < 10^{-19}$ at $5\sigma$ confidence level.

Since the LIV effects are dominated by the most energetic particles, the higher order LIV coefficients can be estimated as
\begin{equation}
  \dhadz R_{\mathrm{cut}}^2 Z^2 = \delta_{\mathrm{had},n} R_{\mathrm{cut}}^{(n+2)} Z^{n+2} \implies \delta_{\mathrm{had},n} = \delta_{\mathrm{had},0} R_{\mathrm{cut}}^{-n} Z^{-n}\; \; ,
\end{equation}
with $Z=1$ taken as the most conservative value. Considering $\dhadz < 10^{-19}$ we have $\dhado < 10^{-38} \ \mathrm{eV}^{-1}$ and $\dhadt < 10^{-57} \ \mathrm{eV}^{-2}$ at $5\sigma$ confidence level for all UHECR scenarios shown in Table~\ref{tab:prop:comb}. These limits are the first obtained with a fit of the spectrum and composition data including UHECR nuclei.

\begin{figure}[!htb]
  \centering
  \begin{tabular}{cc}
    \includegraphics[width=.55\columnwidth]{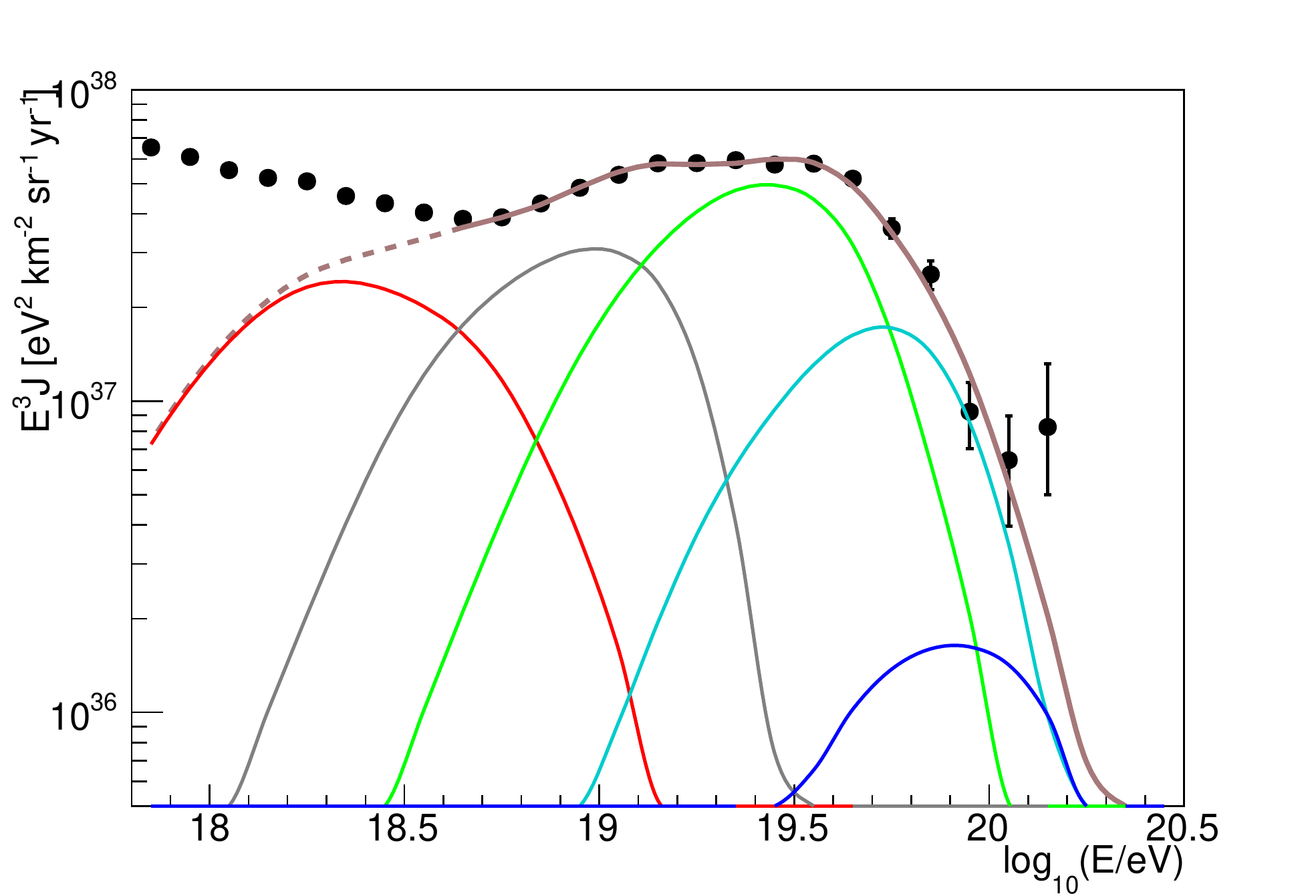}&
    \includegraphics[width=.55\columnwidth]{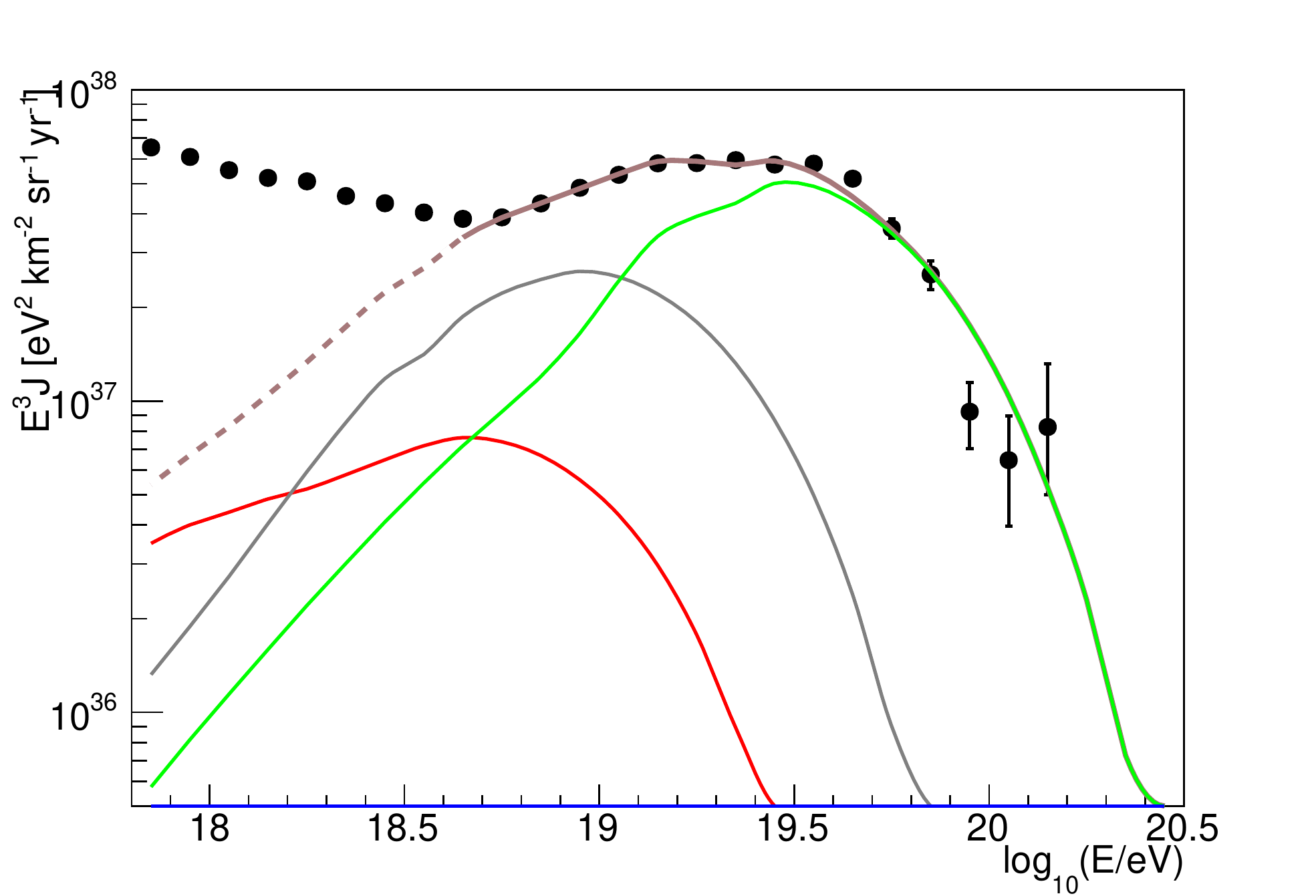}
    \end{tabular}\\
    \centering
    \begin{tabular}{cccc}
   \centering
   \hspace{-0.2cm}
\includegraphics[width=.28\columnwidth]{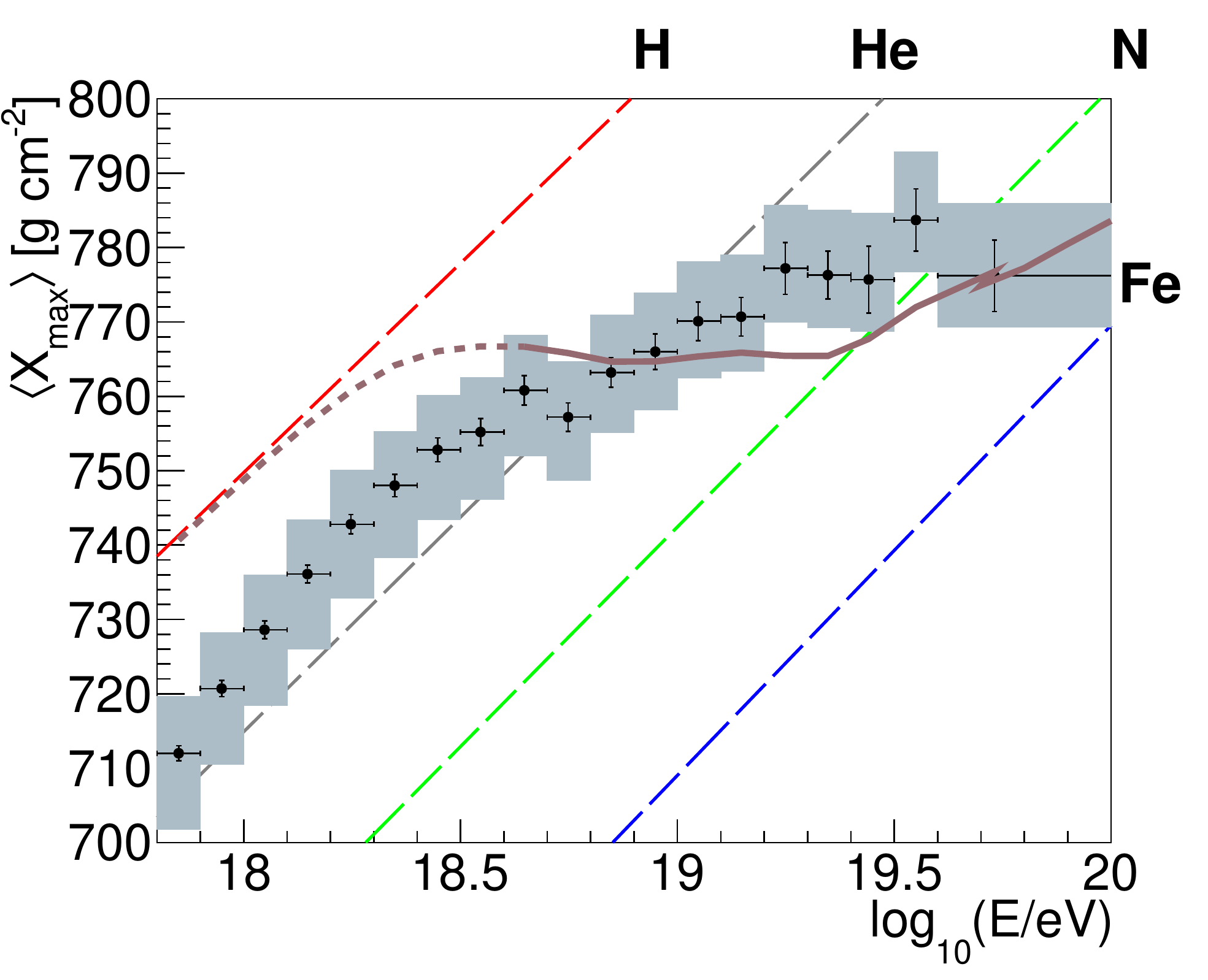}&\hspace{-0.4cm}
\includegraphics[width=.28\columnwidth]{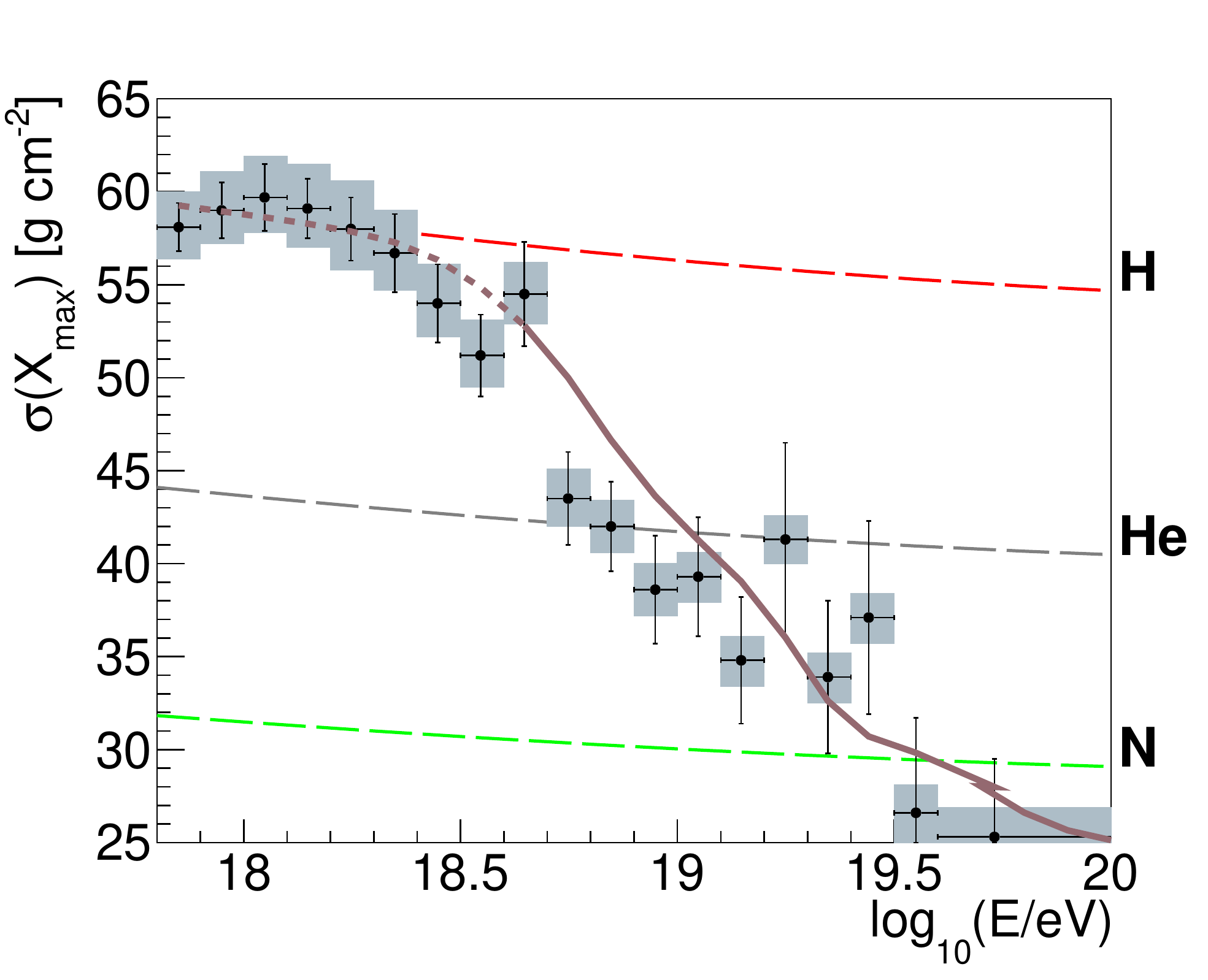}&
\hspace{-0.3cm}
\includegraphics[width=.28\columnwidth]{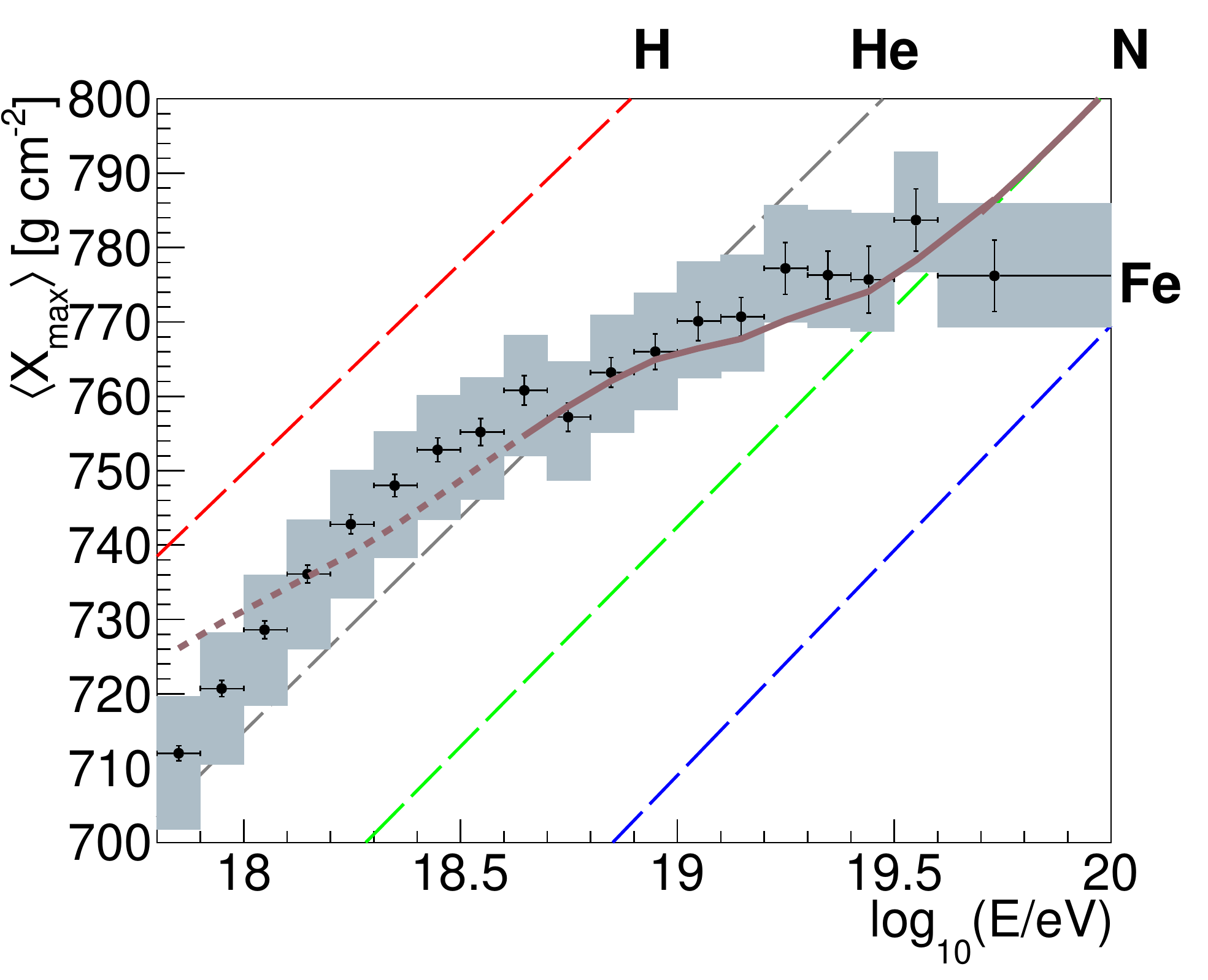}&\hspace{-0.3cm}
\includegraphics[width=.28\columnwidth]{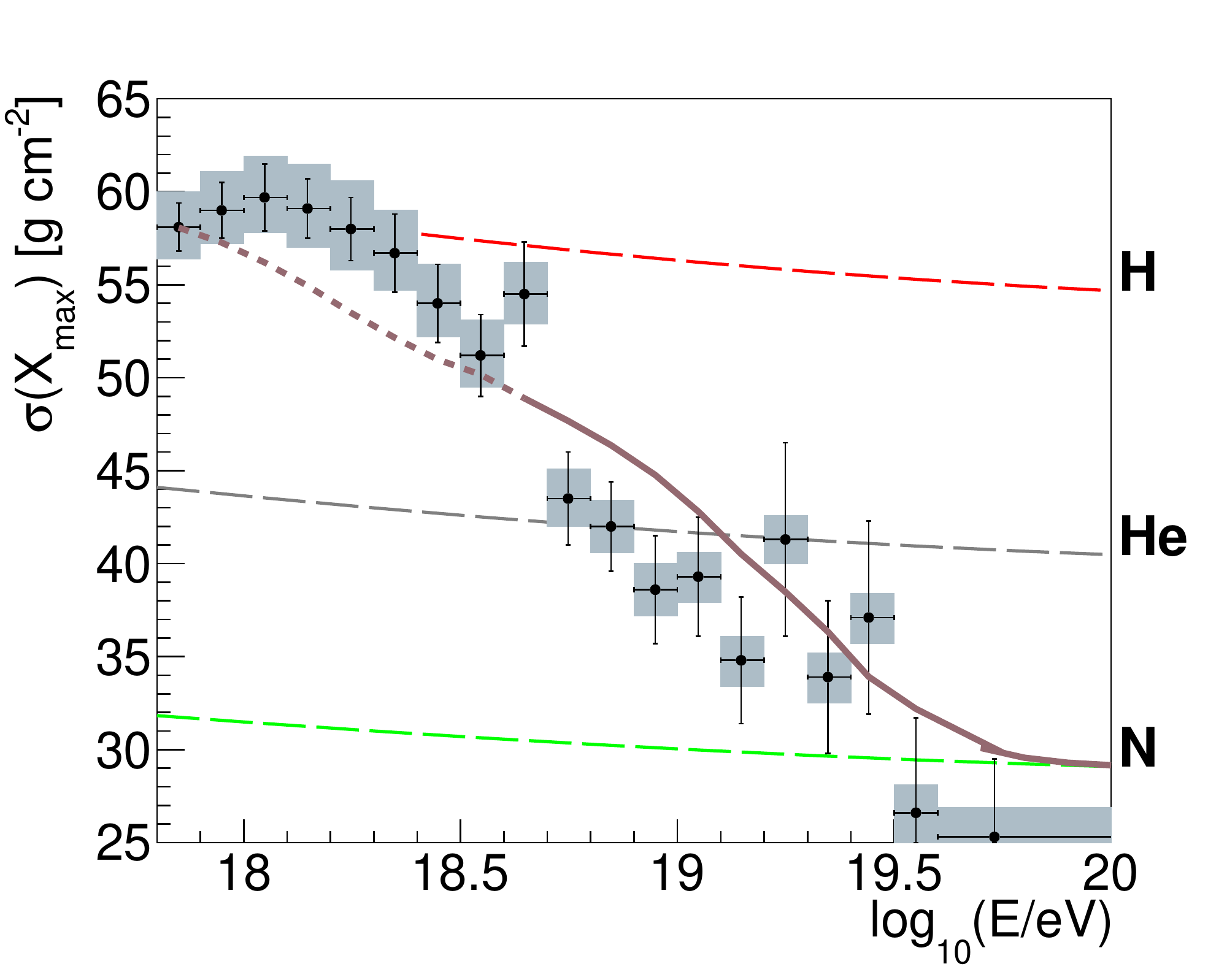}
  \end{tabular}
  \caption {Energy spectrum (top) and first two moments of the \Xmax distributions (bottom), for the LI (left panels) and LIV $\dhadz=10^{-20}$ cases (right panels) for the STGE propagation model compared to the Pierre Auger Observatory data. Partial distributions are grouped according to the mass number as follows: $A=1$ (red), $2 \leq A \leq 4$ (grey), $5 \leq A \leq 22$ (green), $23 \leq A \leq 38$ (cyan), $39 \leq A \leq 56$ (blue), total (brown). Dashed brown lines show the energy region not used for the fit. Dashed lines in the bottom panes show simulations predictions for each element.}
  \label{fig:STGE}
\end{figure}

Table~\ref{tab:bestvalues} shows the value of $\dhadz$ for which we obtained the minimum deviance for each UHECR scenario. Taking this value ($\dhadz = 10^{-21}$) for the STGE model, we show in figure~\ref{fig:STGE} the energy spectrum and the first two moments of the \Xmax distributions in comparison with these corresponding to the LI case. The comparison between the two panels with LI (left panel) and LIV (right panel) shows the correlation between mass composition and LIV in the fit. Lorentz violation suppresses the interactions during propagation which is compensated with a lighter composition at the sources in order to obtain the same composition on Earth.

The effect of the systematic uncertainty in the energy determination of showers measured by the Pierre Auger Observatory was considered. We chose the SPGE scenario for this study and defined the SPGE+14\% and SPGE-14\% scenarios representing the energy shift $\pm 14\%$ within the systematic uncertainties. The entire analysis procedure was repeated considering the energy shift. We verified that the systematic uncertainty of the energy scale has an impact of less than 10\% on the CL of the limits we imposed on the LIV coefficient, $\dhadz$. We have verified that a positive shift of the energy scale would worsen the quality of the fit. This is due to the fact that a positive shift would impose larger rigidity cutoffs or heavier masses at the source, while at the increase of the LIV parameter the interactions are suppressed and thus the preferred scenarios predict lighter masses at the sources. On the other side, corresponding to a negative shift, the same value of the LIV coefficient for which the best description of the data is found does not change.

\begin{table}[]
  \centering
  \begin{tabular}{c|c|c} \hline \hline
    Scenario & $\dhadz$ at minimum deviance & CL \\ \hline
    SPGE & $10^{-21}$ & $5.8\sigma$ \\
    STGE & $10^{-20}$ & $6.2\sigma$ \\
    SPDE & $10^{-20}$ & $5.4\sigma$ \\
    SPGS & $10^{-21}$ & $5.3\sigma$ \\ \hline \hline
  \end{tabular}
  \caption{LIV coefficients $\dhadz$ at minimum deviance and the corresponding confidence level $\sigma = \sqrt{D_{\mathrm{LIV}} - D_{\mathrm{LI}}}$ with respect to the LI case. The model we fit to the data does not describe satisfactorily the measured energy spectrum and \Xmax distributions, thus limiting the interpretation of the minimum deviance as an imprint of LIV in the data as discussed in Section~\ref{sec:disc}.}
  \label{tab:bestvalues}
\end{table}

%==============================================
\section{Discussion and Conclusions}
\label{sec:disc}

In this work, we have explored the use of the data from the Pierre Auger Observatory to test LIV effects. Two independent LIV sectors were explored. The propagation of GZK photons considering subluminal LIV was used to test LIV in the electromagnetic sector by comparison to the measured upper limits on the photon flux. The propagation of UHECR nuclei considering LIV in photopion production and photodesintegration was used to test LIV in the hadronic sector by fitting the measured energy spectrum and \Xmax distributions.

For the electromagnetic sector, we simulated the flux of GZK photons considering LIV in the propagation within several reference and one alternative propagation model. For the reference UHECR scenarios, including the SPGE propagation model which was found to best describe the energy spectrum and \Xmax distribution from the Pierre Auger Observatory under the LI assumption in \cite{Aab:2016zth}, the upper limits on the photons flux imposed by the Pierre Auger Observatory are not constraining enough to set limits on the LIV coefficients. For an alternative scenario with subdominant proton component, limits of $\dgamz > -10^{-21}$, $\dgamo > -10^{-40} \ \mathrm{eV}^{-1}$ and $\dgamt > -10^{-58} \ \mathrm{eV}^{-2}$ are imposed. The limits for such an alternative scenario are several orders of magnitude more constraining than the current most-restrictive limits imposed for subluminal LIV using TeV gamma-ray data~\cite{Lang:2018yog} although the later limits are less dependent on the astrophysical scenario (see~\cite{Martinez-Huerta:2020cut} for a compilation of current limits in the photon sector obtained with astrophysics).

\begin{figure}[h!]
    \centering
    \includegraphics[width=0.7\textwidth]{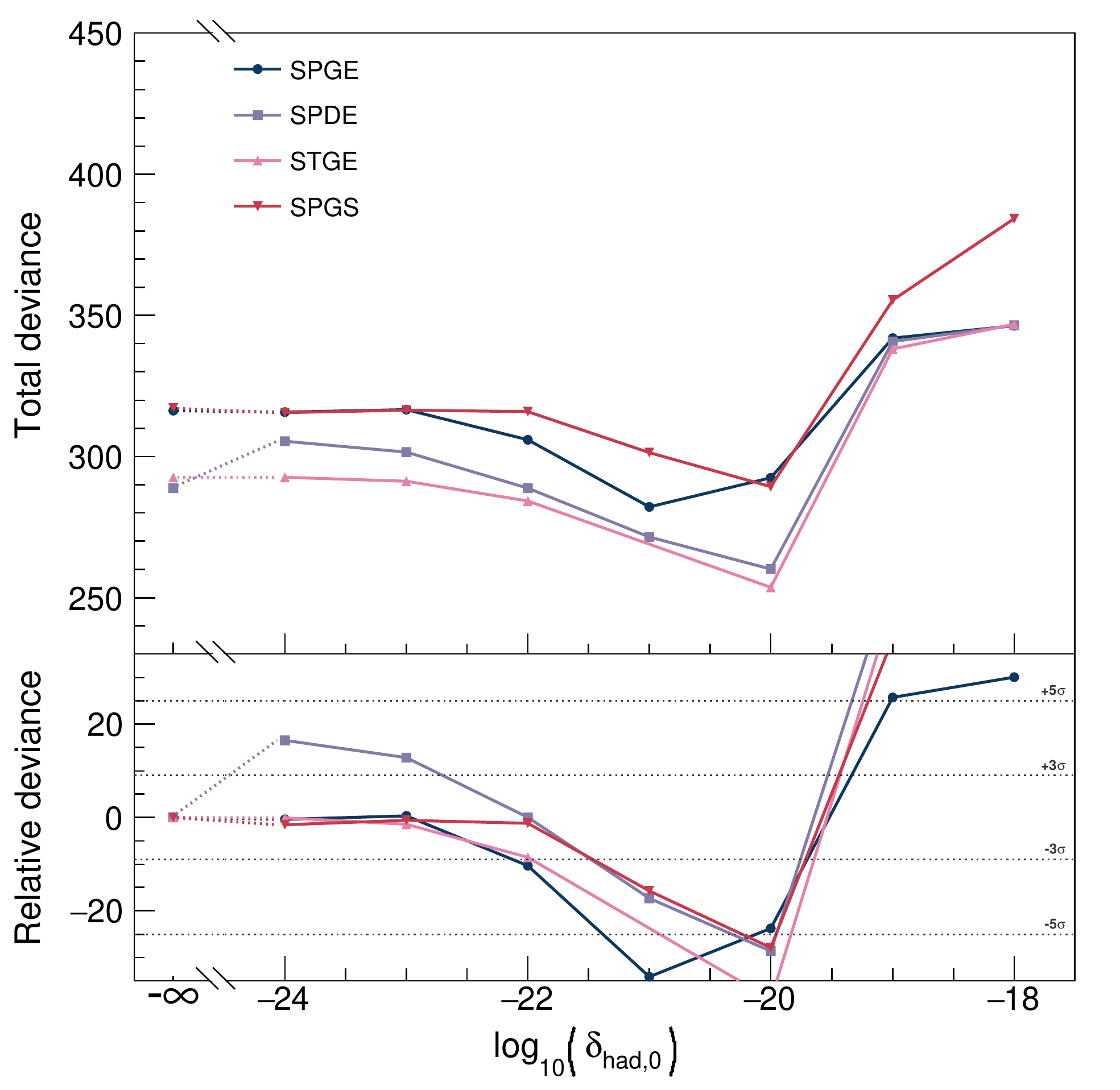}
    \caption{Evolution of the total and relative deviance with respect to the LIV coefficient, $\dhadz$. Each line represents a propagation model. The top panel show the total deviance, while the bottom panel shows the difference in the deviation relative to the LI case. The model we fit to the data does not describe satisfactorily the measured energy spectrum and \Xmax distributions, thus limiting the interpretation of the minimum deviance as an imprint of LIV in the data as discussed in Section~\ref{sec:disc}.}
    \label{fig:deviance}
\end{figure}

In the hadronic sector, the measured energy spectrum and \Xmax distribution were fitted under a LIV assumption for the first time. Limits at $\dhadz < 10^{-19}$, $\dhado < 10^{-38} \ \mathrm{eV}^{-1}$ and $\dhadt < 10^{-57} \ \mathrm{eV}^{-2}$ were imposed at $5\sigma$ CL. These limits are valid for all the UHECR scenarios studied here. The model we fit to the data does not describe satisfactorily the measured energy spectrum and \Xmax distributions. This can be seen by the large total deviance shown in Fig.~\ref{fig:deviance} with values around 300 for 110 data points and 9 free parameters leading to a reduced deviance of about 300/101, thus limiting the interpretation of the minimum deviance between $\dhadz = 10^{-22}$ and $\dhadz = 10^{-20}$ (see Fig.~\ref{fig:deviance}) as an imprint of LIV in the data. The results of the fit showed that the data can be better described by a model with fewer interactions during propagation, but the range of possible UHECR scenarios was not explored as completely as it would be needed for a claim that the data is better described by LIV instead of LI.

In both electromagnetic and hadronic sectors, the predication of the LIV imprints is heavily dependent on the composition of the cosmic rays leaving the source. The upcoming upgrade of the Pierre Auger Observatory, AugerPrime~\cite{AugerPrime2,augerprime3}, will improve our knowledge on the composition of the cosmic rays reaching Earth, and therefore future LIV searches will profit enormously from its operation.

\acknowledgments

\begin{sloppypar}
The successful installation, commissioning, and operation of the Pierre
Auger Observatory would not have been possible without the strong
commitment and effort from the technical and administrative staff in
Malarg\"ue. We are very grateful to the following agencies and
organizations for financial support:
\end{sloppypar}

\begin{sloppypar}
Argentina -- Comisi\'on Nacional de Energ\'\i{}a At\'omica; Agencia Nacional de
Promoci\'on Cient\'\i{}fica y Tecnol\'ogica (ANPCyT); Consejo Nacional de
Investigaciones Cient\'\i{}ficas y T\'ecnicas (CONICET); Gobierno de la
Provincia de Mendoza; Municipalidad de Malarg\"ue; NDM Holdings and Valle
Las Le\~nas; in gratitude for their continuing cooperation over land
access; Australia -- the Australian Research Council; Belgium -- Fonds
de la Recherche Scientifique (FNRS); Research Foundation Flanders (FWO);
Brazil -- Conselho Nacional de Desenvolvimento Cient\'\i{}fico e Tecnol\'ogico
(CNPq); Financiadora de Estudos e Projetos (FINEP); Funda\c{c}\~ao de Amparo \`a
Pesquisa do Estado de Rio de Janeiro (FAPERJ); S\~ao Paulo Research
Foundation (FAPESP) Grants No.~2019/10151-2, No.~2010/07359-6 and
No.~1999/05404-3; Minist\'erio da Ci\^encia, Tecnologia, Inova\c{c}\~oes e
Comunica\c{c}\~oes (MCTIC); Czech Republic -- Grant No.~MSMT CR LTT18004,
LM2015038, LM2018102, CZ.02.1.01/0.0/0.0/16{\textunderscore}013/0001402,
CZ.02.1.01/0.0/0.0/18{\textunderscore}046/0016010 and
CZ.02.1.01/0.0/0.0/17{\textunderscore}049/0008422; France -- Centre de Calcul
IN2P3/CNRS; Centre National de la Recherche Scientifique (CNRS); Conseil
R\'egional Ile-de-France; D\'epartement Physique Nucl\'eaire et Corpusculaire
(PNC-IN2P3/CNRS); D\'epartement Sciences de l'Univers (SDU-INSU/CNRS);
Institut Lagrange de Paris (ILP) Grant No.~LABEX ANR-10-LABX-63 within
the Investissements d'Avenir Programme Grant No.~ANR-11-IDEX-0004-02;
Germany -- Bundesministerium f\"ur Bildung und Forschung (BMBF); Deutsche
Forschungsgemeinschaft (DFG); Finanzministerium Baden-W\"urttemberg;
Helmholtz Alliance for Astroparticle Physics (HAP);
Helmholtz-Gemeinschaft Deutscher Forschungszentren (HGF); Ministerium
f\"ur Innovation, Wissenschaft und Forschung des Landes
Nordrhein-Westfalen; Ministerium f\"ur Wissenschaft, Forschung und Kunst
des Landes Baden-W\"urttemberg; Italy -- Istituto Nazionale di Fisica
Nucleare (INFN); Istituto Nazionale di Astrofisica (INAF); Ministero
dell'Istruzione, dell'Universit\'a e della Ricerca (MIUR); CETEMPS Center
of Excellence; Ministero degli Affari Esteri (MAE); M\'exico -- Consejo
Nacional de Ciencia y Tecnolog\'\i{}a (CONACYT) No.~167733; Universidad
Nacional Aut\'onoma de M\'exico (UNAM); PAPIIT DGAPA-UNAM; The Netherlands
-- Ministry of Education, Culture and Science; Netherlands Organisation
for Scientific Research (NWO); Dutch national e-infrastructure with the
support of SURF Cooperative; Poland -- Ministry of Education and
Science, grant No.~DIR/WK/2018/11; National Science Centre, Grants
No.~2016/22/M/ST9/00198, 2016/23/B/ST9/01635, and 2020/39/B/ST9/01398;
Portugal -- Portuguese national funds and FEDER funds within Programa
Operacional Factores de Competitividade through Funda\c{c}\~ao para a Ci\^encia
e a Tecnologia (COMPETE); Romania -- Ministry of Research, Innovation
and Digitization, CNCS/CCCDI -- UEFISCDI, projects PN19150201/16N/2019,
PN1906010, TE128 and PED289, within PNCDI III; Slovenia -- Slovenian
Research Agency, grants P1-0031, P1-0385, I0-0033, N1-0111; Spain --
Ministerio de Econom\'\i{}a, Industria y Competitividad (FPA2017-85114-P and
PID2019-104676GB-C32), Xunta de Galicia (ED431C 2017/07), Junta de
Andaluc\'\i{}a (SOMM17/6104/UGR, P18-FR-4314) Feder Funds, RENATA Red
Nacional Tem\'atica de Astropart\'\i{}culas (FPA2015-68783-REDT) and Mar\'\i{}a de
Maeztu Unit of Excellence (MDM-2016-0692); USA -- Department of Energy,
Contracts No.~DE-AC02-07CH11359, No.~DE-FR02-04ER41300,
No.~DE-FG02-99ER41107 and No.~DE-SC0011689; National Science Foundation,
Grant No.~0450696; The Grainger Foundation; Marie Curie-IRSES/EPLANET;
European Particle Physics Latin American Network; and UNESCO.
\end{sloppypar}

\bibliography{main}
\bibliographystyle{unsrt}

%==============================================
\newpage
%figures

\end{document}